\documentclass[prb,twocolumn,showpacs,floatfix]{revtex4-1}

\usepackage[utf8]{inputenc}
\usepackage{graphicx}
\usepackage{flafter}
\usepackage{amsmath}
\usepackage{amssymb}
\usepackage{float}
\usepackage{color}
\usepackage{subfigure}
\definecolor{dg}{rgb}{0.00, 0.40, 0.29}

\let \k \relax
\newcommand{\k}{{\bf k}}
\newcommand{\q}{{\bf q}}
\newcommand{\Q}{{\bf Q}}
\newcommand{\pdagger}{{\phantom{\dagger}}}
\newcommand{\beps}{\bar\varepsilon}

\begin{document}

\title{Fate of the excitonic insulator in the presence of phonons}
\author{B. Zenker$^1$, H. Fehske$^1$, and H. Beck$^2$}
\affiliation{$^1$Institut f{\"u}r Physik,
                  Ernst-Moritz-Arndt-Universit{\"a}t Greifswald,
                  D-17489 Greifswald, Germany \\
                  $^2$D{\'e}partement de Physique and Fribourg Center for Nanomaterials, 
                  Universit{\'e} de Fribourg,
                  CH-1700 Fribourg, Switzerland}
\date{\today}
\begin{abstract}
The influence of phonons on the formation of the excitonic insulator has hardly been analyzed so far. Recent experiments on $\rm Ta_2NiSe_5$, 1$T$-$\rm TiSe_2$, and $\rm TmSe_{0.45}Te_{0.55}$, being candidates for realizing the excitonic-insulator state, suggest, however, that the underlying lattice plays a significant role. Employing the Kadanoff-Baym approach we address this issue theoretically. We show that owing  to the electron-phonon coupling  a static lattice distortion may arise at the excitonic instability. Most importantly such a distortion  will destroy the acoustic phase mode being present if the electron-hole pairing and condensation is exclusively driven by the Coulomb interaction. The absence of off-diagonal long-range order, when lattice degrees of freedom are involved, challenges that excitons in these materials form a superfluid condensate of Bose particles or Cooper pairs composed of electrons and holes.
\end{abstract}
\maketitle

\section{Introduction}
\label{sec:Intro}
The excitonic insulator (EI) is a longstanding problem in condensed matter physics. Although first theoretical work dates back  almost half a century,~\cite{Mo61, Kno63, KK64, *KK65, JRK67, HR68a} the experimental realization of the EI phase has proven to be quite challenging. In recent years a number of  mixed-valent rare-earth chalcogenide and transition-metal dichalcogenide materials have been investigated,~\cite{BSW91, CMCBDGBAPBF07, WSTMANTKNT09} which are promising in this respect and have renewed the interest in the EI also from the theoretical side.~\cite{BF06, IPBBF08, MCCBSDGBABFP09, ZIBF10, PBF10, ZIBF12}

In particular, the  mechanism of the formation of the EI has been analyzed in detail.~\cite{BF06,IPBBF08,PBF10,SEO11,ZIBF12,EKOF14} In the weak coupling, semimetallic regime the Coulomb-driven EI formation reveals a formal analogy to the BCS theory of superconductivity. In the strong coupling, semiconducting regime, on the other hand, the transition to the anticipated EI phase is a Bose-Einstein condensation (BEC) of preformed excitons. Then, within the EI, a smooth crossover from a BCS- to a BEC-like state should occur.

An EI instability  can  be triggered by the Coulomb interaction between electrons and holes. Therefore, the  theoretical modeling typically focuses on a purely electronic mechanism. First attempts to include a coupling to the lattice degrees of freedom have been made quite recently, motivated by several experiments indicating that the lattice is involved at the phase transition to the anticipated EI phase.~\cite{MBCAB11,KTKO13,KTKO13err,ZFBMB13,PBF13}
For example, in the TmSe$_{0.45}$Te$_{0.55}$ compound  a drop of the specific heat and an increase of the lattice constant have been interpreted as a strong coupling between excitons and phonons.~\cite{WBM04}
Furthermore, in 1$T$-TiSe$_2$ there is a longstanding debate whether the charge-density wave and the concomitant structural phase transition observed in this material are the results of an excitonic~\cite{CMCBDGBAPBF07, MCCBSDGBABFP09} or a lattice instability.~\cite{Hu77,RKS02} A combination of both instabilities was also proposed.~\cite{KMCC02,WNS10} Without any doubt, lattice effects are crucial in this material.
Finally, at the transition to the suggested EI phase in Ta$_2$NiSe$_5$ the lattice structure changes from orthorhombic to monoclinic, although the charge does not modulate.~\cite{SCFWDSI86,KTKO13,KTKO13err} Therefore, the electron-phonon interaction seems non-negligible in this material as well.

Motivated by these findings, we analyze the EI formation in the framework of a rather generic two-band model that comprises both the Coulomb interaction and an explicit electron-phonon coupling. Besides its relevance to the materials under study, some fundamental theoretical questions are brought up in this model. So we address the electron-hole pair spectrum and the nature of the ordered ground state.

The paper is organized as follows. In Sec.~\ref{sec:model} we introduce our model. 
A mean-field treatment in terms of the electron Green functions is given in Sec.~\ref{sec:meanfield}. 
In Sec.~\ref{sec:selfEnergy} we calculate the electronic self-energies using a Kadanoff-Baym approach. From this, we argue that the considered electron-phonon interaction does not lead to a qualitative modification of the single-particle spectra.
The electron-hole pair spectrum, on the other hand, indicates a strong influence of the  phonons. This is shown in Sec.~\ref{sec:ehpair}. How the lattice dynamics affects the electron-hole pairing is analyzed in the framework of the Kadanoff-Baym scheme. We present some numerical results in Sec.~\ref{sec:results} and show that the purely electronic model possesses an acoustic mode, whereas the collective mode becomes massive if phonons participate.
In Sec.~\ref{sec:condensate} we discuss the problem of off-diagonal long-range order.
A short summary of our results is given in Sec.~\ref{sec:conclusion}.

\section{Model}
\label{sec:model}
For our analysis, we start from a two-band model with interband Coulomb interaction and an explicit electron-phonon coupling,
\begin{equation}
H=H_{\rm e} + H_{\rm e-e} + H_{\rm ph} + H_{\rm e-ph}.
\label{H}
\end{equation}
The noninteracting band-electron contribution is given by
\begin{equation}
H_{\rm e} = \sum_\k \varepsilon_{\k v} c_{\k v}^\dagger c_{\k v}^\pdagger
+ \sum_\k \varepsilon_{\k c} c_{\k c}^\dagger c_{\k c}^\pdagger \,,
\label{H_e}
\end{equation}
where $c_{\k\sigma}^{(\dagger)}$ is the annihilation (creation) operator for an electron with momentum $\k$ in the valence band (band index $\sigma=v$) or in the conduction band ($\sigma=c$).  
The corresponding band dispersions are denoted as $\varepsilon_{\k\sigma}$.
We consider a valence band (conduction band) with a single, nondegenerate maximum (minimum). Moreover, the electron-electron interaction is supposed to be
\begin{equation}
H_{\rm e-e} = \sum_{\k,\k',\q} \frac{V(\q)}{N} c_{\k c}^\dagger c_{\k+\q c}^\pdagger 
c_{\k' v}^\dagger c_{\k'-\q v}^\pdagger ,
\label{H_ee}
\end{equation}
where $V(\q)$ is the effective Coulomb repulsion. $N$ is the number of unit cells.
In harmonic approximation, the phonon Hamiltonian reads
\begin{equation}
H_{\rm ph} = \sum_\q \omega_\q b_\q^\dagger b_\q^\pdagger \,,
\label{H_ph}
\end{equation}
where $\omega_\q$ is the bare phonon frequency, and $b_\q^{(\dagger)}$ is the annihilation (creation) operator for a phonon with momentum $\q$.  Throughout this paper we set $\hbar=1$.

If the electron-phonon interaction is assumed to be
\begin{eqnarray}
H_{\rm e-ph} &=& \sum_{\k,\q} \bigg( \frac{g_{-\q}}{\sqrt{N}}
(b_{-\q}^\dagger +b_\q^\pdagger) c_{\k+\q c}^\dagger c_{\k v}^\pdagger 
\nonumber \\
&&+ \frac{g_\q}{\sqrt{N}} (b_\q^\dagger + b_{-\q}^\pdagger ) c_{\k v}^\dagger c_{\k+\q c}^\pdagger \bigg) ,
\label{H_eph}
\end{eqnarray}
the phonon directly couples to an electron-hole pair with the (real) coupling constant $g_\q$. 
Then, the annihilation of a phonon is inevitably connected with a transfer of an electron from the valence band to the conduction band and vice versa. Such a coupling of phonons to excitons may look rather specific, but for materials near the semimetal-semiconductor transition (SM-SC) it is of relevance.

In order to model the SM-SC transition, we consider the case of half-filling,
\begin{equation}
n_c + n_v = 1,
\label{halfFilling}
\end{equation}
where $n_\sigma = \frac{1}{N} \sum_\k \langle c_{\k\sigma}^\dagger c_{\k\sigma}^\pdagger \rangle$.

\section{Mean-field Green functions}
\label{sec:meanfield}
The electron-phonon coupling~\eqref{H_eph} may cause a deformation of the lattice at sufficiently low temperatures.\cite{MMAB12b}
A static lattice distortion is characterized by
\begin{equation}
\delta_{\bar\Q} = \frac{2}{\sqrt{N}} g_{\bar\Q} \langle b_{\bar\Q}^\dagger \rangle ,
\label{phOP}
\end{equation}
where the ordering vector of the dimerized phase is denoted as $\bar\Q$. Working at half-filling, we assume that $\bar\Q$ is either zero or half a reciprocal lattice vector. Then $b_{\bar\Q}^\dagger=b_{-\bar\Q}^\dagger$. 
As a consequence, the parameter $\delta_{\bar\Q}$ is a real number that measures the amplitude of the static lattice distortion. 
For charge-density-wave systems with more complex lattice deformations, e.g., the chiral charge-density wave in the transition metal-dichalcogenide 1$T$-TiSe$_2$, $\delta_{\bar\Q}$ might be complex.\cite{ZFBMB13} Nevertheless, since $\delta_{\bar\Q}^\ast = \delta_{-\bar\Q}$, the static lattice distortion---in real space---is a real quantity.
Adopting the frozen phonon approximation,  we replace the phonon operators by their averages. Then, the Hamiltonian~\eqref{H} describes an effective electronic system.

Applying subsequently a Hartree-Fock decoupling scheme, our model reduces to
\begin{align}
H^{\rm MF} =& \sum_\k \beps_{\k v} c_{\k v}^\dagger c_{\k v}^\pdagger 
+\sum_\k \beps_{\k+\bar\Q c} \, c_{\k+\bar\Q c}^\dagger c_{\k+\bar\Q c}^\pdagger
\nonumber \\
&+\sum_\k \left( x_{\k \bar\Q} c_{\k+\bar\Q c}^\dagger c_{\k v}^\pdagger 
+ x_{\k \bar\Q}^\ast c_{\k v}^\dagger c_{\k+\bar\Q c}^\pdagger \right) ,
+C_{\rm dec} 
\label{H_MF}
\end{align}
with  renormalized dispersions $\beps_{\k \sigma} = \varepsilon_{\k \sigma} + V(0) n_{-\sigma}$.
In Eq.~\eqref{H_MF}, 
\begin{equation}
x_{\k\bar\Q} = \delta_{\bar\Q} - \Delta_{\k\bar\Q} 
\label{x_kQ}
\end{equation}
is the gap parameter, 
\begin{equation}
\Delta_{\k\bar\Q} = \frac{1}{N}\sum_{\k'} V(\k'-\k+\bar\Q) 
\langle c_{\k' v}^\dagger c_{\k'+\bar\Q c}^\pdagger \rangle
\label{Delta_kQ}
\end{equation}
is the Coulomb-induced hybridization between the valence band and the conduction band, and  
\begin{eqnarray}
C_{\rm dec} &=& \frac{1}{N} \sum_{\k,\k'} V(\k'-\k+\bar\Q) 
\langle c_{\k+\bar\Q c}^\dagger c_{\k v}^\pdagger \rangle 
\langle c_{\k' v}^\dagger c_{\k'+\bar\Q c}^\pdagger \rangle
\nonumber \\
&& + \frac{N}{4} \frac{\omega_{\bar\Q}}{|g_{\bar\Q}|^2} \delta_{\bar\Q}^2 
- N V(0) n_c n_v .
\label{MF_decouplingC}
\end{eqnarray}
For an undistorted lattice $\Delta_{\k\bar\Q}$ serves as the EI order parameter, whose phase is undetermined and can be chosen arbitrarily.~\cite{ZFB10,ZFBMB13} 
A finite electron-phonon interaction removes this freedom.

The gap equation that determines $\Delta_{\k\bar\Q}$ and the conservation of the particle number [Eq.~\eqref{halfFilling}] are valid on both sides of the SM-SC transition, i.e., these relations hold in the BCS as well as BEC regimes.\cite{ZIBF12}

In mean-field approximation the electronic Green functions become 
\begin{eqnarray}
G_v(\k, z_1) &=&\langle\langle c_{\k v}^\pdagger ; c_{\k v}^\dagger \rangle \rangle
\nonumber \\
&=& v_\k^2 G_A(\k,z_1) + u_\k^2 G_B(\k,z_1) , 
\label{Gv_HF}  
\end{eqnarray}
\begin{eqnarray}
G_c(\k+\bar\Q, z_1) &=& \langle\langle c_{\k+\bar\Q c}^\pdagger ; 
c_{\k+\bar\Q c}^\dagger \rangle\rangle 
\nonumber \\
&=& u_\k^2 G_A(\k,z_1) + v_\k^2 G_B(\k,z_1) , 
\label{Gc_HF} 
\end{eqnarray}
\begin{eqnarray}
F(\k,z_1) &=& \langle\langle c_{\k+\bar\Q c}^\pdagger ; c_{\k v}^\dagger \rangle \rangle 
\nonumber \\
&=& -u_\k v_\k \big[ G_B(\k,z_1)-G_A(\k,z_1) \big]
\nonumber \\
&=& \langle\langle c_{\k v}^\pdagger ; c_{\k+\bar\Q c}^\dagger \rangle\rangle
=F^\dagger(\k,z_1) ,
\label{F_HF}
\end{eqnarray}
where $z_1$ denotes fermionic Matsubara frequencies, and
\begin{equation}
G_{A/B}(\k,z_1) = \frac{1}{z_1-E_{\k A/B}}\;,
\label{GaB_HF}
\end{equation}
\begin{eqnarray}
E_{\k A/B} &=& \frac{1}{2}(\beps_{\k+\bar\Q c}+\beps_{\k v}) 
\nonumber \\
&&\pm \sqrt{\frac{1}{4} (\beps_{\k+\bar\Q c}-\beps_{\k v})^2 +
|x_{\k \bar\Q}| ^2}  \;,
\label{quasEn_HF}
\end{eqnarray}
\begin{equation}
u_\k^2 / v_\k^2 =  \frac{1}{2}\pm \frac{\frac{1}{4}(\beps_{\k+\bar\Q c}
-\beps_{\k v})}{\sqrt{\frac{1}{4} (\beps_{\k+\bar\Q c}-\beps_{\k v})^2 + 
|x_{\k\bar\Q}|^2}}\;.
\label{cohFac_1}
\end{equation}

One can easily show that $|\Delta_{\k\bar\Q}| \propto |\delta_{\bar\Q}|$.~\cite{ZFBMB13}
Moreover, $\delta_{\bar\Q}$ and $\Delta_{\k\bar\Q}$ couple to the same set of operators and, therefore, enter the quasiparticle dispersion in an equal manner. Hence, at the mean-field level of approximation we cannot discriminate between a Coulomb-driven or a phonon-driven phase transition.

\section{Electronic self-energy}
\label{sec:selfEnergy}
We now analyze self-energy effects.
To this end, we use the technique developed by Kadanoff and Baym and determine  the self-energy of the electrons.~\cite{KB62} 
The imaginary-time Green functions are defined as
\begin{eqnarray}
G_v(\k,t-t') &=& -i\langle T[c_{\k v}^\pdagger(t) c_{\k v}^\dagger(t')] \rangle ,
\label{Gv_time} \\
G_c(\k,t-t') &=& -i\langle T[c_{\k c}^\pdagger(t) c_{\k c}^\dagger(t')] \rangle ,
\label{Gc_time} \\
F(\k,t-t') &=& -i\langle T[c_{\k+\bar\Q c}^\pdagger(t) c_{\k v}^\dagger(t')] \rangle ,
\label{F_time} \\
F^\dagger(\k,t-t') &=&-i\langle T[c_{\k v}^\pdagger(t) c_{\k+\bar\Q c}^\dagger(t')]\rangle,
\label{Fdagger_time} 
\end{eqnarray}
with imaginary-time variables $t$ and $t'$.

We start from the equation of motion (EOM) for the valence-electron Green function,
\begin{align}
\bigg( i\frac{\partial}{\partial t}-\varepsilon_{\k v} \bigg) & G_v(\k,t-t')
=\delta(t-t') 
\nonumber \\
&- i\sum_\q \frac{g_\q}{\sqrt{N}} G_2^P(\k,\q,t,t')
\nonumber \\
&-i\sum_{\k',\q} \frac{V_c(\q)}{N} G_2^V(\k,\k',\q,t,t') ,
\label{EOM_Gv}
\end{align}
where
\begin{equation}
G_2^V(\k,\k',\q,t,t') = \left\langle T[c_{\k-\q v}^\pdagger(t) c_{\k' c}^\dagger(t)
c_{\k'+\q c}^\pdagger(t) c_{\k v}^\dagger(t') ] \right\rangle ,
\label{G2V} 
\end{equation}
\begin{equation}
G_2^P(\k,\q,t,t') = \left\langle  T[(b_\q^\dagger(t)+b_{-\q}^\pdagger(t)) 
c_{\k+\q c}^\pdagger(t) c_{\k v}^\dagger(t') ] \right\rangle ,
\label{G2P}
\end{equation}
and proceed as follows: The auxiliary correlation functions~\eqref{G2V} and \eqref{G2P}  are expanded up to first order in the interactions they couple to, i.e.,  $G_2^V(\k,\k',\q,t,t')$ is expanded up to linear order in $V_c(\q)$, and $G_2^P(\k,\q,t,t')$ is expanded up to linear order in $g_\q$. Subsequently, we decouple the correlation functions taking only  electron-hole fluctuations into account.

Straight forward calculation yields
\begin{align}
\bigg( i\frac{\partial}{\partial t}-\beps_{\k v} \bigg)& G_v(\k,t-t')
=\delta(t-t') + x_{\k\bar\Q} F(\k,t-t') 
\nonumber \\
&-\int_0^{-i\beta} d\!\tau\, \sigma_{vv}(\k,t-\tau) G_v(\k,\tau-t')
\nonumber \\
&- \int_0^{-i\beta} d\!\tau\, \sigma_{vF}(\k,t-\tau) F(\k,\tau-t') 
\label{EOM_Gv_final}
\end{align}
(here $\beta$ is the inverse temperature), 
with the self-energies
\begin{align}
\sigma_{vv}(\k,t&-\tau) = \frac{1}{N^2} \sum_{\q,\q',\Q} V_c(\q) V_c(\q') G_c(\k+\Q,t-\tau)
\nonumber \\
&\times G_2(\Q,\k+\q',\k-\q,\tau-t)
\nonumber \\
&-\frac{i}{N}\sum_\q |g_\q|^2 D(\q,\tau-t) G_c(\k+\q,t-\tau) ,
\label{sigma_vv} 
\end{align}
\begin{align}
\sigma_{vF}&(\k,t-\tau) = \frac{1}{N^2}\sum_{\q,\q',\Q} V_c(\q) V_c(\q') F(\k+\Q+\bar\Q,t-\tau)
\nonumber \\
&\times F_2(\Q,\k+\bar\Q-\q'+\Q,\k-\q,\tau-t)
\nonumber \\
&-\frac{i}{N}\sum_\q |g_\q|^2 D(\q,\tau-t) F(\k+\q+\bar\Q,t-\tau) .
\label{sigma_vF}
\end{align}

The electron-hole pair correlation functions are defined as
\begin{equation}
G_2(\Q,\k,\k',t-t') = -\left\langle T[c_{\k v}^\dagger(t) c_{\k+\Q c}^\pdagger(t)
c_{\k'+\Q c}^\dagger(t') c_{\k' v}^\pdagger(t') ]\right\rangle ,
\label{G2_1} 
\end{equation}
\begin{equation}
F_2(\Q,\k,\k',t-t') = -\left\langle T[c_{\k-\Q c}^\dagger(t) c_{\k v}^\pdagger(t)
c_{\k'+\Q c}^\dagger(t') c_{\k' v}^\pdagger(t') ]\right\rangle .
\label{F2_1}
\end{equation}
The phonon Green function is given by
\begin{equation}
D(\q,t-t') = -i\left\langle T[(b_{-\q}^\dagger (t)+b_\q^\pdagger (t)) (b_\q^\dagger (t') + b_{-\q}^\pdagger (t')) ]\right\rangle .
\label{D}
\end{equation}

With the same procedure we obtain the EOM of the conduction-electron Green function and the EOM of the anomalous Green function. These equations can be found in  Appendix~\ref{app:EOMs}.

Note that both the electron-electron interaction and the electron-phonon interaction couple different species  (valence electrons, conduction electrons, and electrons in the hybridized state) to each other. 
The structure of the self-energies shows that the one-particle spectrum cannot be used--at least at this level of approximation--to decide whether the ordered ground state is the effect of the Coulomb interaction alone or if phonons contribute. Let us therefore analyze the electron-hole pair spectrum in the following.

\section{Electron-hole pair spectrum}
\label{sec:ehpair}
In the Bethe-Salpeter equation, describing the correlations of electron-hole pairs, the Coulomb interaction is treated in ladder approximation.~\cite{KKER86} In the vicinity of the SM-SC transition, the small number of free electrons and holes makes two-particle collisions to be the dominant process. The ladder approximation takes the sequence of these collisions into account and is suitable to describe both the build-up of  excitons and the formation of the EI.~\cite{ZIBF12}

We now work out the influence of $H_{\rm e-ph}$ [Eq.~\eqref{H_eph}] on the electron-hole pairs.
The four-time electron-hole pair correlation functions are defined as
\begin{align}
G_2(\Q & ,\k,\k',t_1,t_2,t_3,t_4) =
\nonumber \\
& -\left\langle T[ c_{\k v}^\dagger(t_1) c_{\k+\Q c}^\pdagger(t_2) 
c_{\k'+\Q c}^\dagger(t_4) c_{\k' v}^\pdagger (t_3) ]\right\rangle,
\label{G2_fourtime}
\end{align}
\begin{align}
F_2(\Q & ,\k,\k',t_1,t_2,t_3,t_4) = 
\nonumber \\
&-\left\langle T[ c_{\k-\Q c}^\dagger(t_1)c_{\k v}^\pdagger(t_2) 
c_{\k'+\Q c}^\dagger (t_4) c_{\k' v}^\pdagger(t_3) ]\right\rangle.
\label{F2_fourtime}
\end{align}
The relations to the two-time electron-hole pair correlation functions, occurring in Sec.~\ref{sec:selfEnergy}, are $G_2(\Q,\k,\k',t-t') = G_2(\Q,\k,\k',t,t,t',t')$ and $F_2(\Q,\k,\k',t-t') = F_2(\Q,\k,\k',t,t,t',t')$.
In order to analyze the effects of the phonons within the Kadanoff-Baym scheme,~\cite{KB62} we expand the correlation functions~\eqref{G2_fourtime} and \eqref{F2_fourtime}  to  leading order in  the electron-phonon coupling.  
Restricting ourselves to the study of electron-hole pairs, there are no incoming or outgoing phonon branches. Hence, the phonons must be created and annihilated in one diagram, and the first non-vanishing contribution is of second order in the electron-phonon coupling constant  $g_\q$. 
The many-particle correlation functions that occur in the leading-order expansion of $G_2(\Q,\k,\k',t_1,t_2,t_3,t_4)$ and $F_2(\Q,\k,\k',t_1,t_2,t_3,t_4)$  are subsequently decoupled  into electron-hole pair correlation functions, electron Green functions, and phonon Green functions. We identify two effects of $H_{\rm e-ph}$: Excitons can be created (annihilated) by the annihilation (creation) of a phonon, and phonons may change the individual momenta of the electron and the hole in the bound state without modifying the momentum of the exciton. This is illustrated by the diagrams depicted in Fig.~\ref{fig:diagrams}. 
The explicit equations for the electron-hole pair correlation functions can be found in  Appendix~\ref{app:correlationFuncs}.

\begin{figure}[h]
\centering
\subfigure{\includegraphics[width=0.3\linewidth]{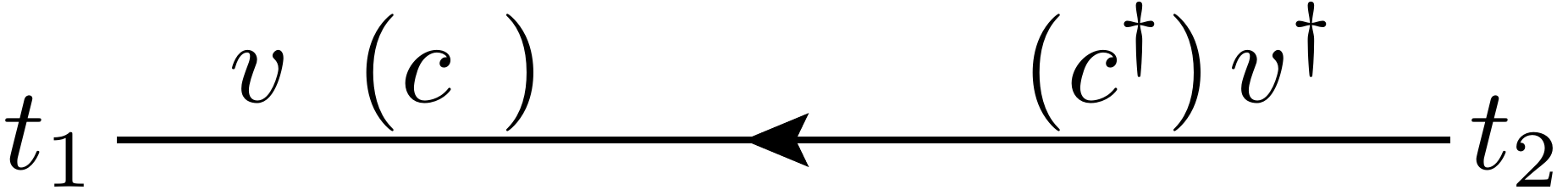}}
\hspace{0.2\linewidth}
\subfigure{\includegraphics[width=0.3\linewidth]{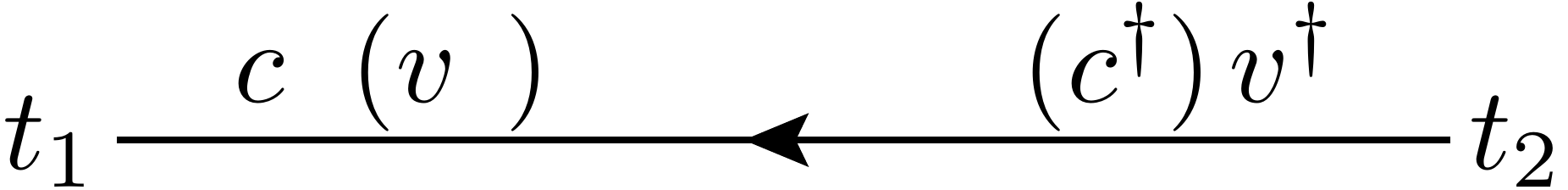}}
\\
\vspace{0.6cm}
\subfigure{\includegraphics[width=0.49\linewidth]{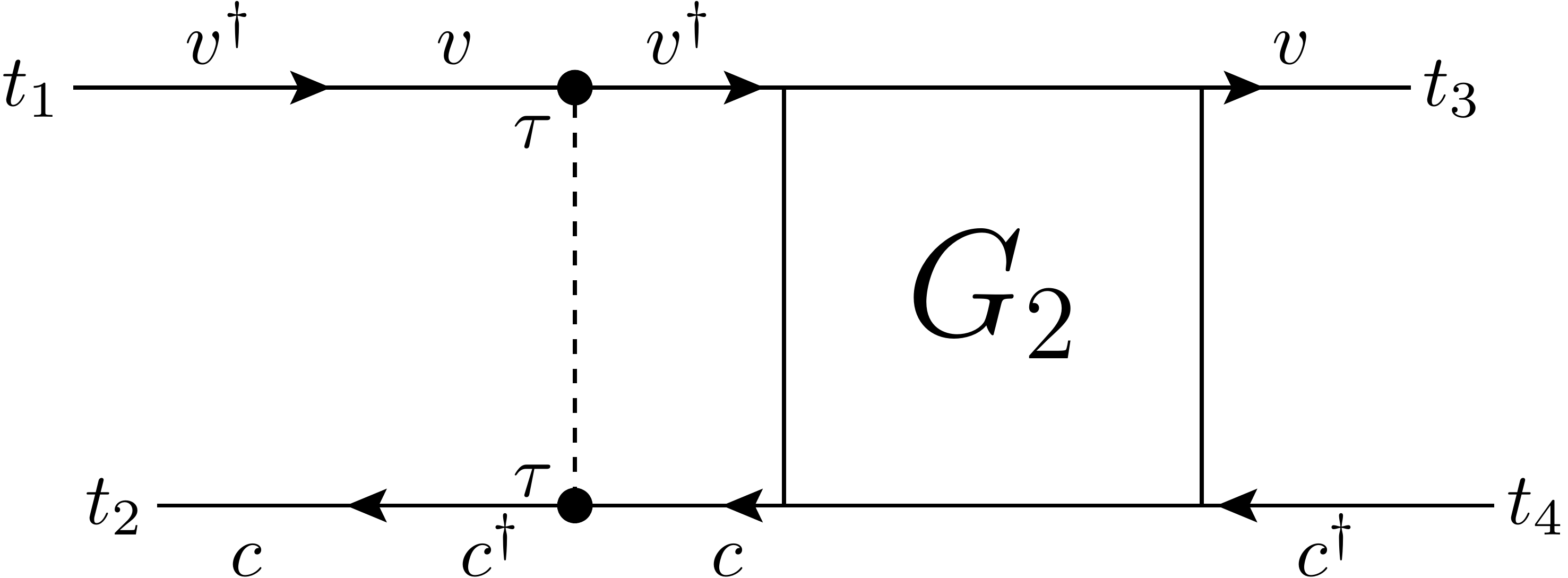}}
\hfill
\subfigure{\includegraphics[width=0.49\linewidth]{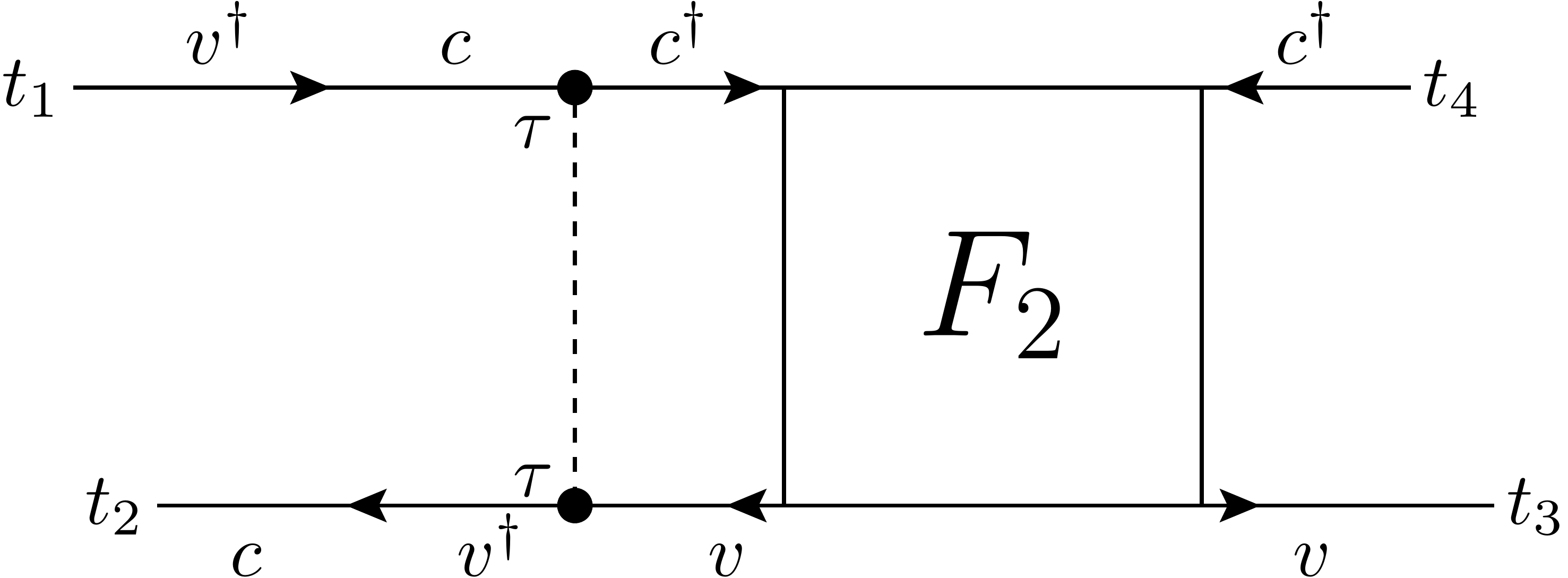}}
\\
\subfigure{\includegraphics[width=0.49\linewidth]{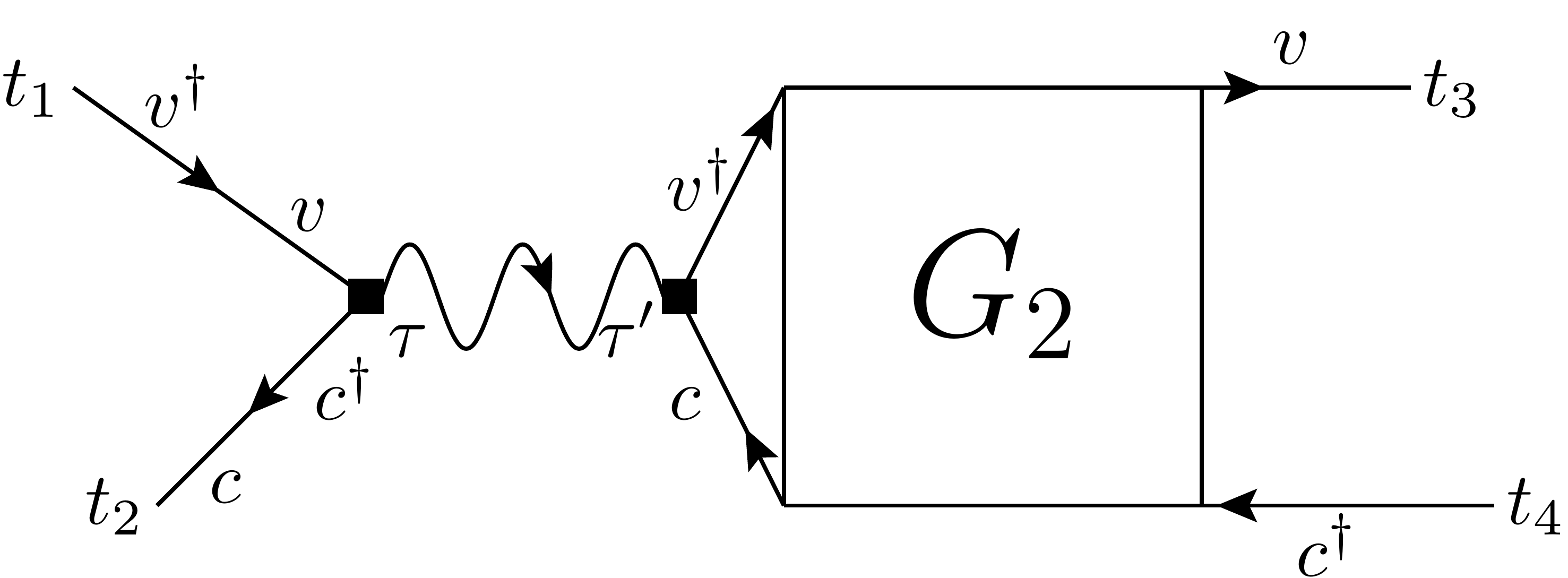}}
\hfill
\subfigure{\includegraphics[width=0.49\linewidth]{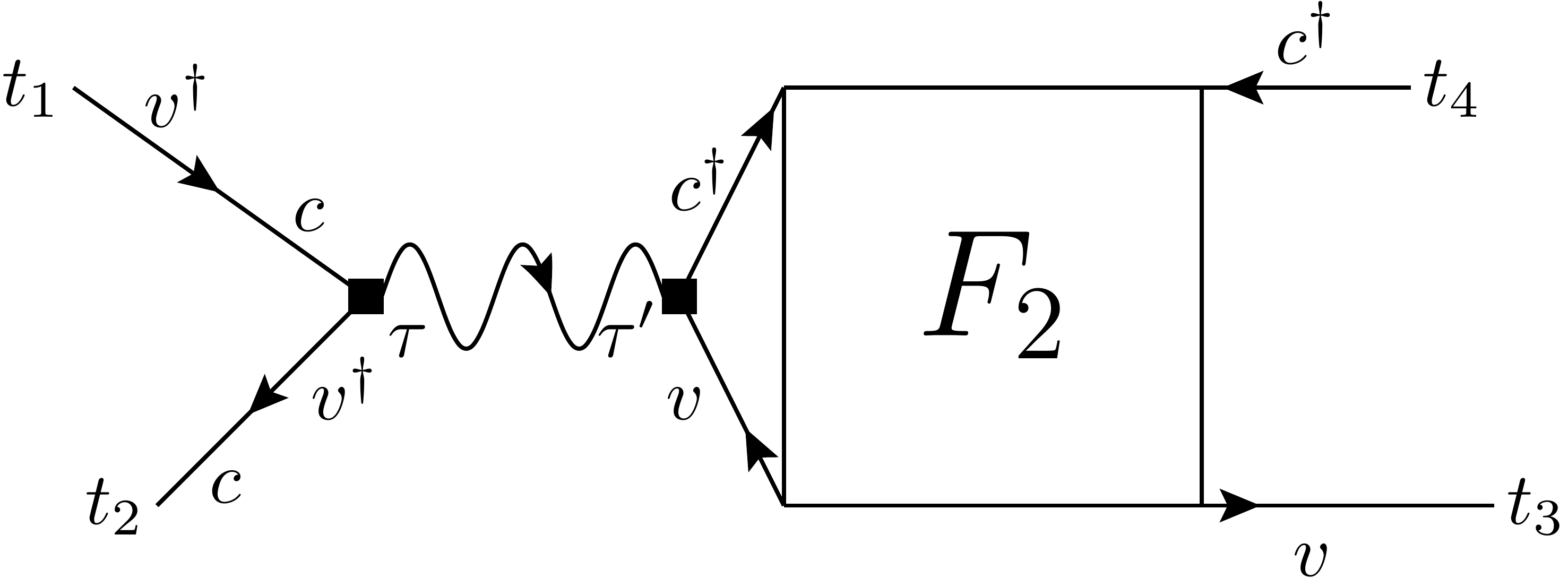}}
\\
\subfigure{\includegraphics[width=0.49\linewidth]{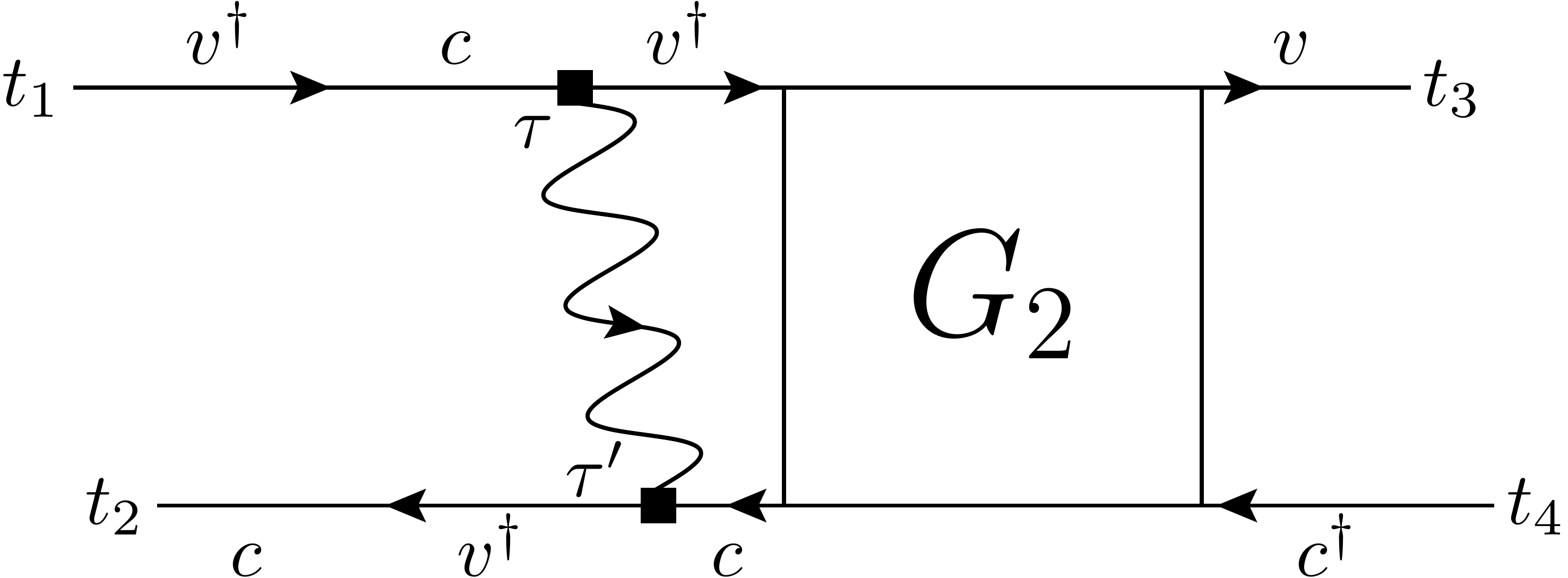}}
\hfill
\subfigure{\includegraphics[width=0.49\linewidth]{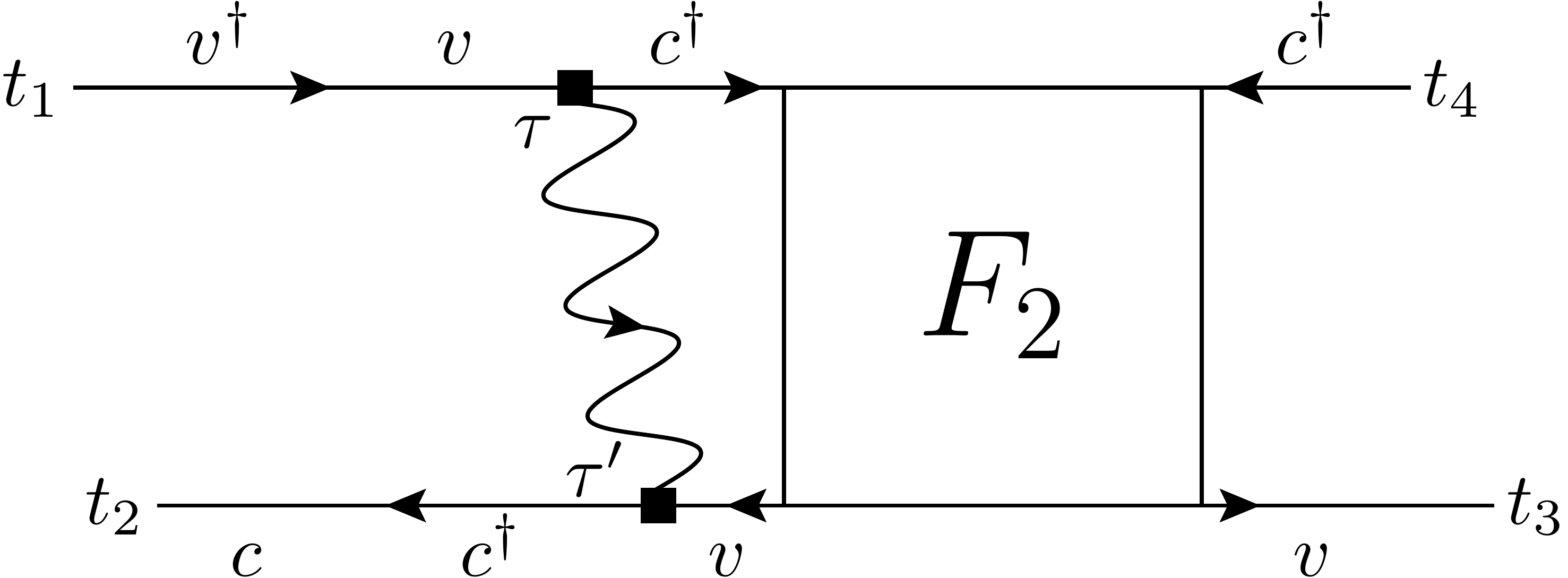}}
\caption{Diagrams occurring in the equations for the electron-hole pair correlation functions. First row: Single-particle Green functions $G_v$ [$G_c$] (left-hand side) and $F$ [$F^\dagger$] (right-hand side). Second row: Ladder approximation for the Coulomb interaction. Third row: Ring diagrams including the electron-phonon interaction. Fourth row: Ladder diagrams including  the electron-phonon interaction. The dashed lines with the vertex points represent the Coulomb interaction, the wavy lines represent the phonon Green function, and the vertex squares represent our electron-phonon interaction.}
\label{fig:diagrams}
\end{figure}

Following Ref.~\onlinecite{JRK67}, we analyze the collective modes by finding poles of the ``phase" correlation function
\begin{equation}
P(\Q,z_\nu) = X(\Q,z_\nu) - Y(\Q,z_\nu) ,
\label{Phase_func}
\end{equation}
where 
\begin{align}
X(\Q,z_\nu) =& \left( \frac{1}{-i\beta}\right)^{2} 
\frac{i}{N} \sum_{\k,\k'}  \sum_{z_{2}, z_{3}}
G_2(\Q,\k,\k',z_\nu-z_{2},z_{2},z_{3}),
\label{G2_oneTime} \\
Y(\Q,z_\nu) =& \left( \frac{1}{-i\beta}\right)^{2} 
\frac{i}{N} \sum_{\k,\k'}  \sum_{z_{2}, z_{3}}
F_2(\Q,\k,\k',z_\nu-z_{2},z_{2},z_{3}).
\label{F2_oneTime}
\end{align}

\section{Results and Discussion}
\label{sec:results}
\begin{figure*}
\centering
\includegraphics[width=0.4\linewidth]{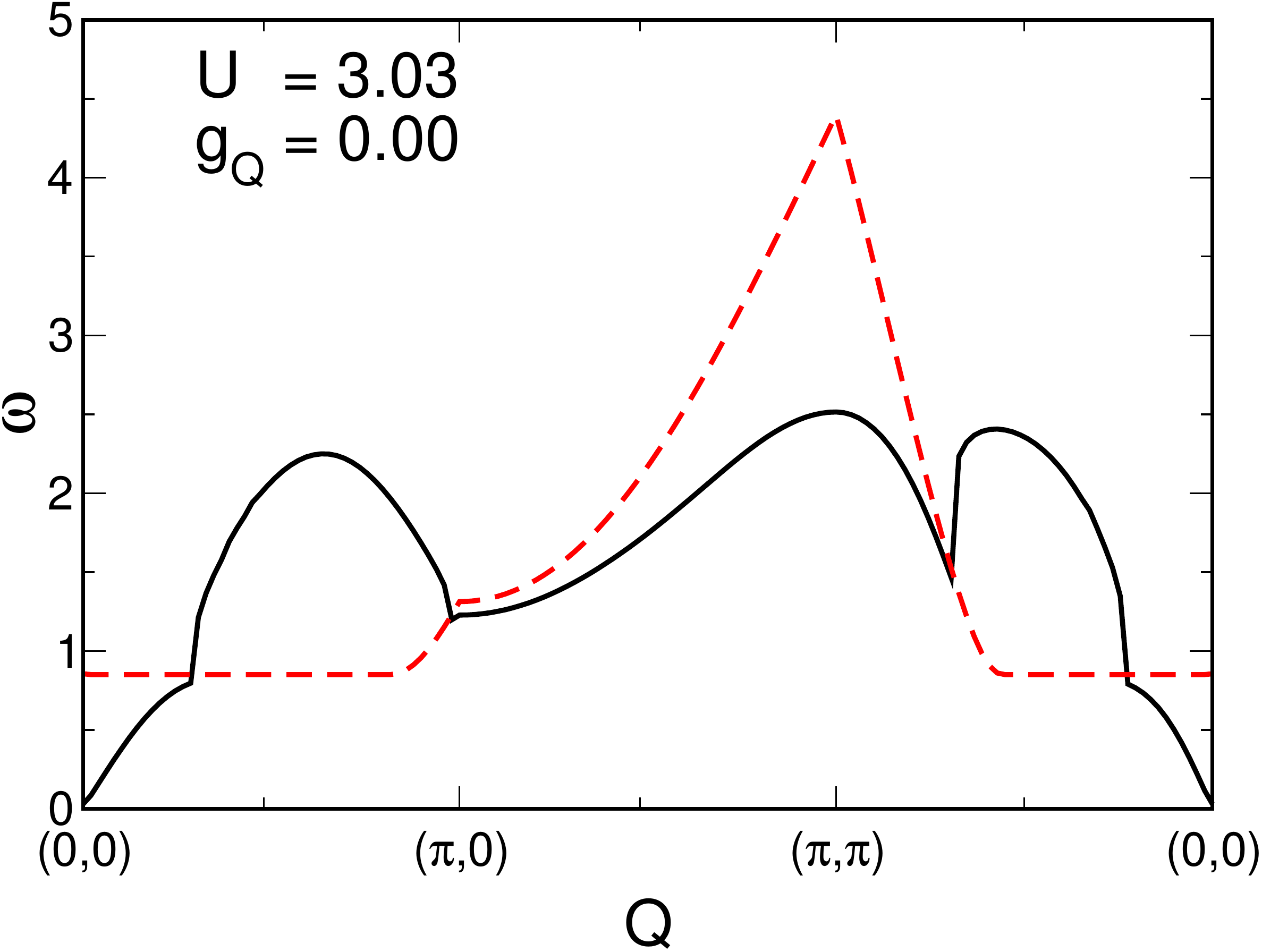}
\hspace{0.1\linewidth}
\includegraphics[width=0.4\linewidth]{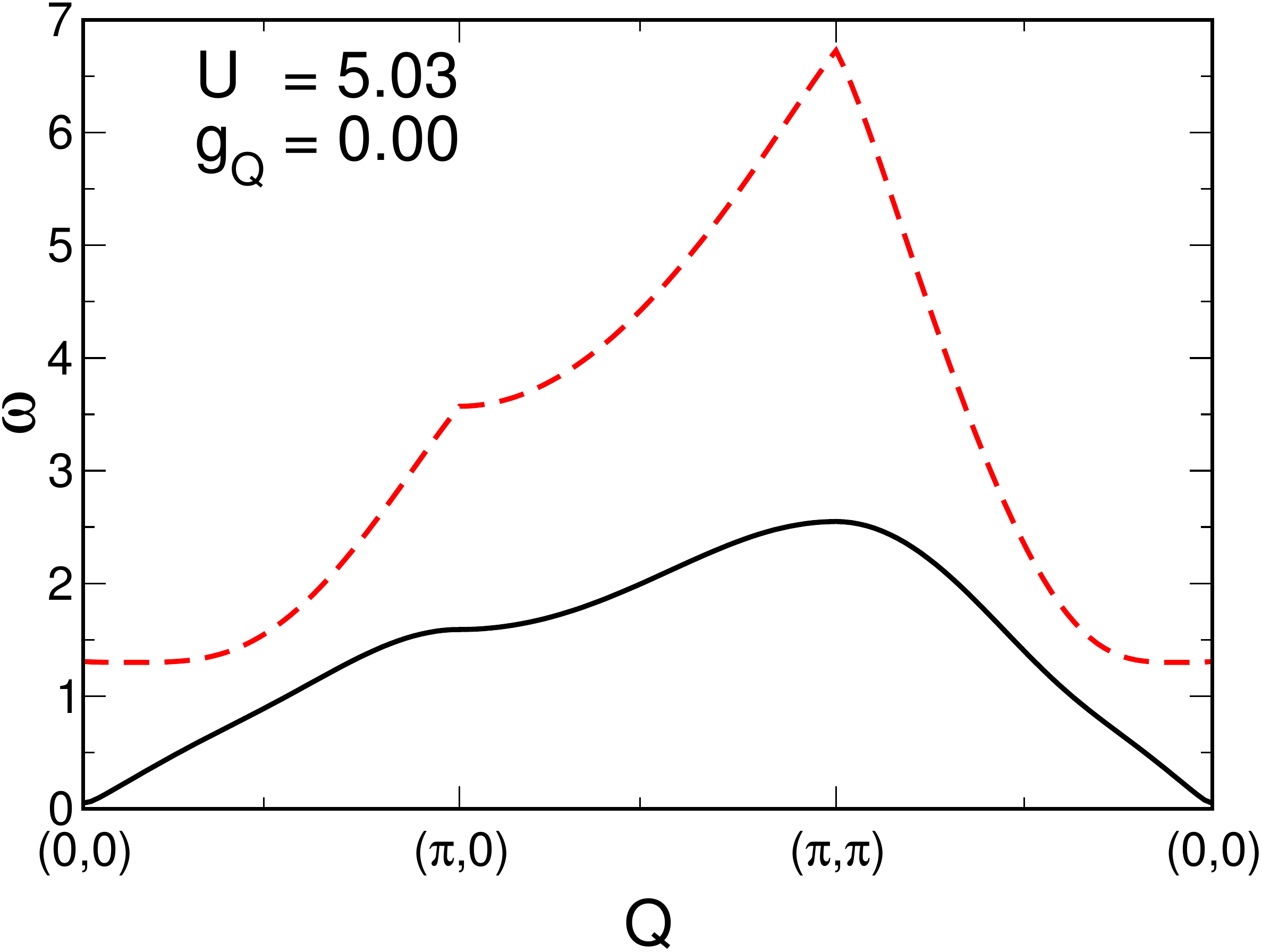}
\caption{(Color online) Electron-hole excitation spectrum at zero temperature  without electron-phonon coupling. Black  solid lines show the phase mode, red dashed lines display the lower boundary of the electron-hole continuum. Results are given for the BCS-type pairing regime with  $U=3.03$ (left panel) and  the BEC-type pairing regime with $U=5.03$ (right panel).}
\label{fig:modes_EI}
\end{figure*}
In the numerical  evaluation of the equations derived so far we work at zero temperature and assume a local Coulomb potential [$V(\q)=U$], a momentum-independent electron-phonon coupling ($g_{\q}=g_{\bar\Q}$), and dispersionless Einstein phonons ($\omega_\q = \omega_{\bar\Q}$). We furthermore consider a direct band-gap situation, i.e., the valence-band maximum and the conduction-band minimum are located at the Brillouin-zone center. Then, the ordering vector of the low-temperature phase is $\bar\Q=0$. 
For $\bar\Q\neq 0$ the EI with lattice deformation is accompanied by a charge-density wave. Apart from that, the situation for a finite ordering vector corresponds to the situation considered here.

To avoid hard numerics, we consider a two-dimensional (square) lattice. For this, the  bare band dispersions $\varepsilon_{\k\sigma}= E_\sigma -2t_\sigma [\cos(k_x)+\cos(k_y)]$ ($\sigma=v,c$), where $t_c$ sets the unit of energy.   Typical  model parameters are: $E_v=-2.4$, $E_c=0$,  $t_v=-0.8$, and $\omega_{\bar\Q}=0.01$. 
Let us emphasize that the present analytical calculations and the scenario that will be discussed below hold for both  a bare semimetallic and semiconductive band structure. Furthermore, the two-dimensional (square) lattice is used for the sake of convenience only, the  results obtained below stay valid also for three-dimensional systems (and in this case also for finite temperatures).
Performing  the analytic continuation $z_\nu\rightarrow \omega+i\delta$ we take $\delta=2\cdot 10^{-3}$. Moreover, we utilize the Hartree-Fock single-particle Green functions in the calculation.

\subsection{Vanishing electron-phonon coupling}
\label{sec:Phasemode}
We start our analysis for a system, where the phonons are neglected  ($g_{\bar\Q}=0$). In this case, the correlation function~\eqref{Phase_func} can  be calculated according to
\begin{align}
P(\Q,z_\nu) &=\frac{ X^{(0)}(\Q,z_\nu)\left[ 1+ a(-\Q,-z_\nu)+b(\Q,z_\nu)\right]} {L(\Q,z_\nu)}
\nonumber \\
&-\frac{Y^{(0)}(\Q,z_\nu) \left[1+a(\Q,z_\nu)+b(\Q,z_\nu)\right]}{ L(\Q,z_\nu)} ,
\label{calc_P}
\end{align}
where
\begin{align}
a(\Q,z_\nu) =& UX^{(0)}(\Q,z_\nu) , 
\label{a} \\
b(\Q,z_\nu) =& UY^{(0)}(\Q,z_\nu) ,
\label{b}
\end{align}
and the denominator reads
\begin{align}
L(\Q,z_\nu) =& \left[1+a(\Q,z_\nu)\right] \left[1+a(-\Q,-z_\nu)\right]
-\left[b(\Q,z_\nu)\right]^2 .
\label{denom}
\end{align}
The definitions of $X^{(0)}$ and $Y^{(0)}$ are analogous to Eqs.~\eqref{G2_oneTime} and \eqref{F2_oneTime}, respectively, except that we use the [according to Eq.~\eqref{FourierTrafo} transformed] bare electron-hole pair correlation functions \eqref{G2_0} and \eqref{F2_0}. 

Figure~\ref{fig:modes_EI} shows the so-called ``phase mode" for weak and strong couplings.  Obviously, there exists a gapless phase mode in the EI state, i.e., $\omega(\Q)\rightarrow 0$ for $\Q\rightarrow 0$.~\cite{KM65a, *KM65b, KM65c, *KM66a, JRK67, LEKMSS04}  The appearance of this mode can be attributed to the $U(1)$ symmetry of the underlying electronic model $H=H_{\rm e}+H_{\rm e-e}$.~\cite{La66b} Because of  this symmetry the phase of $\Delta_{\k\bar\Q}$ can be chosen arbitrarily, which results in such an acoustic mode.

\begin{figure*}
\centering
\includegraphics[width=0.4\linewidth]{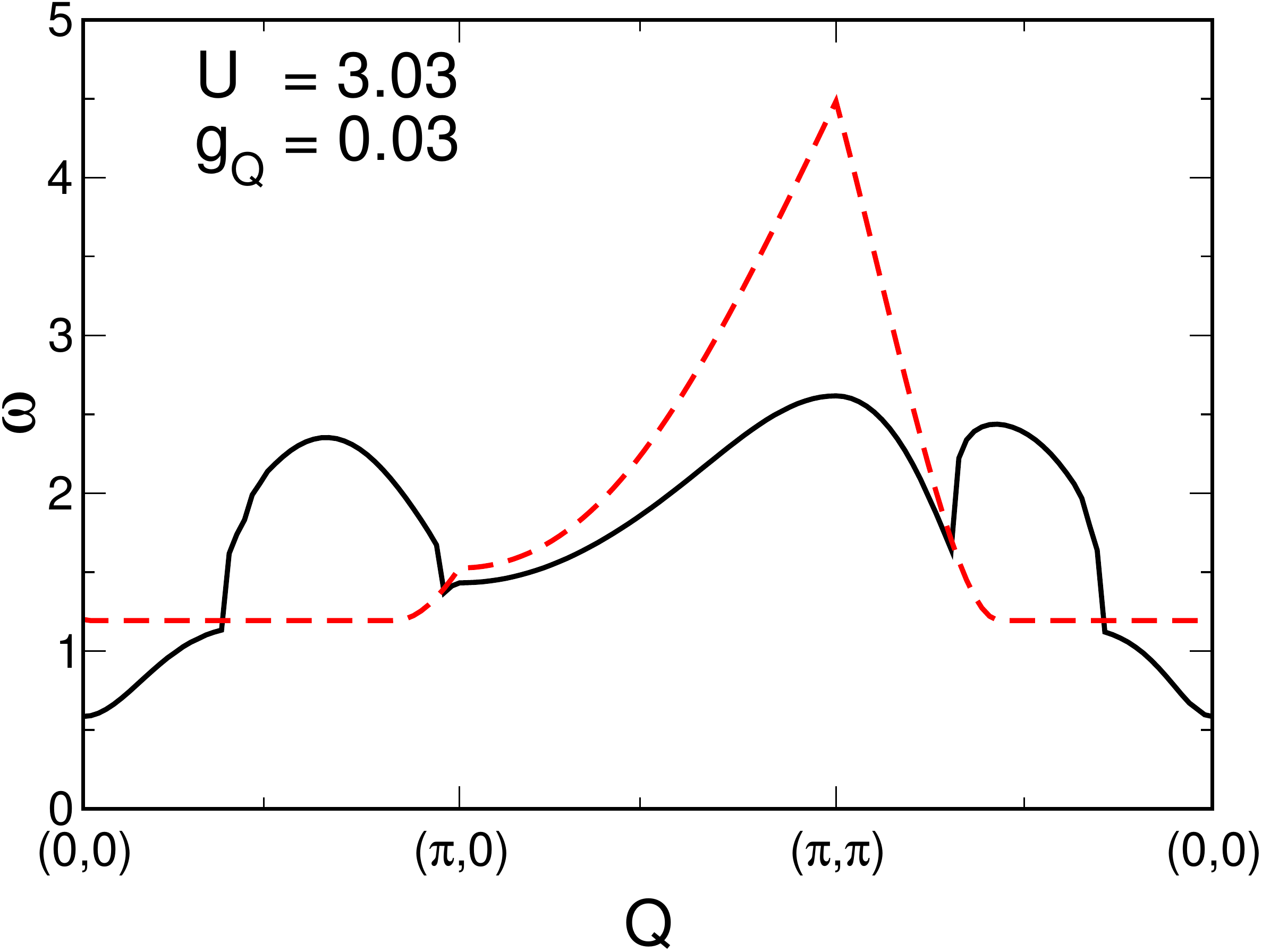}
\hspace{0.1\linewidth}
\includegraphics[width=0.4\linewidth]{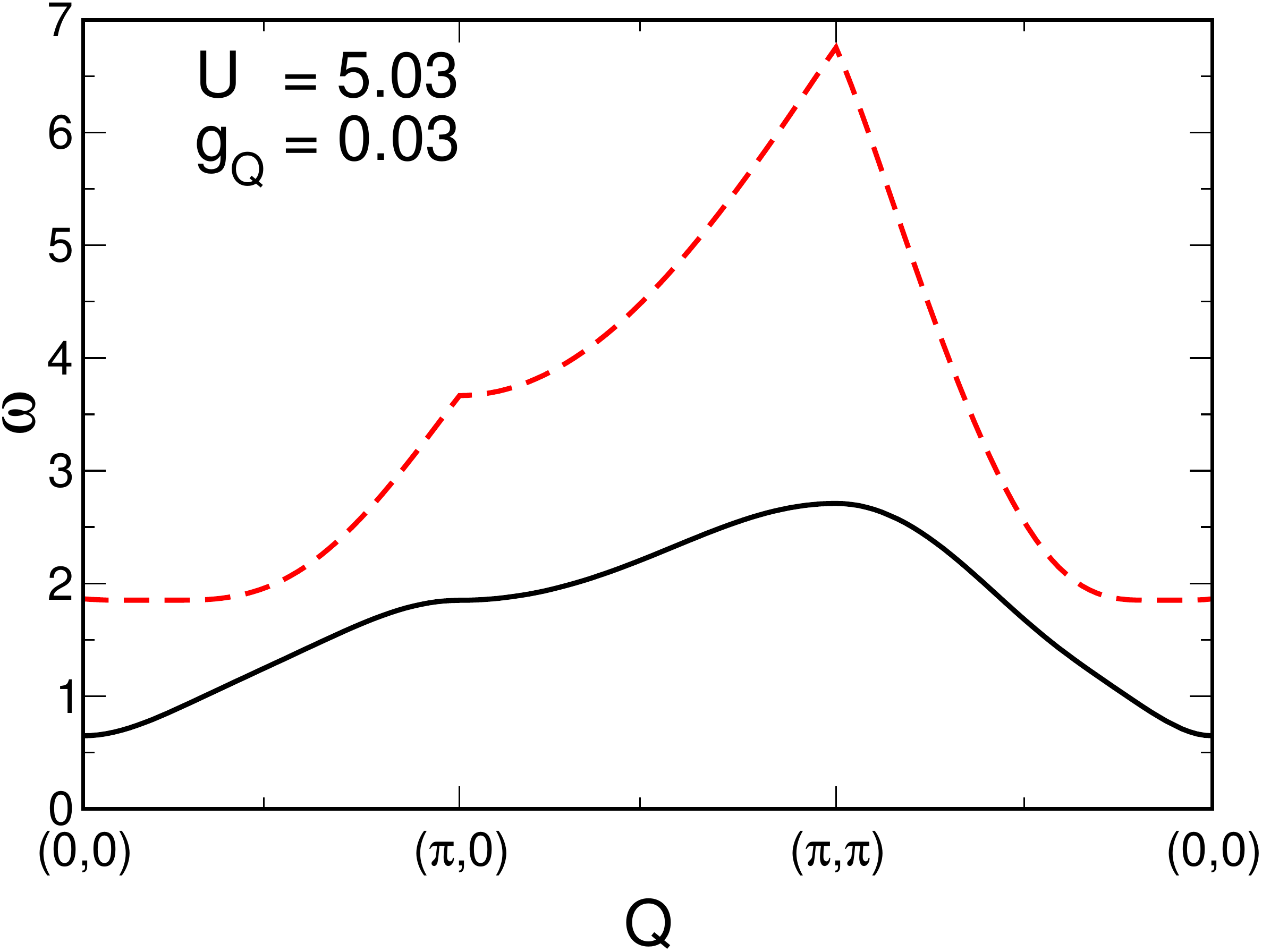}
\caption{(Color online) Electron-hole excitation spectrum for  a distorted lattice at zero temperature. The black, solid lines show the phase mode and the red, dashed lines show the lower boundary of the electron-hole continuum.}
\label{fig:modes_dist}
\end{figure*}

Figure~\ref{fig:modes_EI} furthermore reveals the different character of the phase mode for weak- and strong-coupling situations. In the weak-coupling, BCS-type pairing regime ($U=3.03$) $\omega(\Q)$ exhibits a steep increase for small momenta and, as a result, quickly enters the electron-hole continuum, which it leaves again close to the Brillouin-zone corner. The lower boundary of the electron-hole continuum is given by
\begin{equation}
\omega_C(\Q) = {\rm min}_{\k} (E_{\k+\Q A}-E_{\k B}) ,
\label{continuum}
\end{equation}
where $E_{\k A}$ and $E_{\k B}$ ($E_{\k A}>E_{\k B}$) are the renormalized quasiparticle energies in the ordered ground state. In Hartree-Fock approximation  $E_{\k A,B}$ follow from  Eq.~\eqref{quasEn_HF}.
The momentum dependence of the excitation energy of the mode changes remarkably when the boundary to the electron-hole continuum is crossed. Contrariwise, in the strong-coupling, BEC-type pairing regime, the collective phase mode entirely lies below the electron-hole continuum and is a smooth function.~\cite{LEKMSS04}

The existence of an acoustic phase mode can be understood as follows. Here, the static uniform limit of the noninteracting phase correlation function is well defined, i.e., 
\begin{align}
&\lim_{\omega\rightarrow 0} \left[ \lim_{\Q\rightarrow 0} P^{(0)}(\pm\Q,\pm\omega) \right]
= \lim_{\Q\rightarrow 0} \left[ \lim_{\omega\rightarrow 0} P^{(0)}(\pm\Q,\pm\omega) \right]
\nonumber \\
&= \lim_{\omega,\Q\rightarrow 0}P^{(0)}(\pm\Q,\pm\omega)  = P^{(0)}(0,0) .
\label{SUlimit1}
\end{align}
According to Eq.~\eqref{SUlimit1} and since we consider only  interband correlations, the static, uniform limit of $P(\Q,\omega)$ exists, contrary to the case when additional intraband correlations are taken into account.~\cite{LP14}
We find for the static, uniform phase correlation function 
\begin{equation}
P(0,0) = \frac{P^{(0)}(0,0)}{ 1+U P^{(0)}(0,0) } .
\label{PhasecorrSceneA}
\end{equation}
The  (Hartree-Fock) gap equation~\eqref{x_kQ} is 
\begin{equation}
1+U P^{(0)}(0,0)=0 .
\label{GapEqSceneA}
\end{equation}
Comparing Eq.~\eqref{PhasecorrSceneA} with Eq.~\eqref{GapEqSceneA} unveils that $P(0,0)$ exhibits a pole; hence, the phase mode is acoustic.

\subsection{Static electron-phonon coupling}
\label{sec:dist_Coulomb}
Let us now discuss the behavior of the phase mode if the lattice deforms at the EI phase transition, i.e., we have $\delta_{\bar\Q}\neq 0$.
According to the strong coupling of electron-hole pair fluctuations and phonons, the phonon frequency is significantly renormalized at the SM-SC transition and might even vanish at low temperatures, leading to a static deformation of the lattice.~\cite{MMAB12b}

The lattice distortion is contained in the electron Green functions but does not explicitly appear in the Bethe-Salpeter equation. Hence, the phase correlation function is determined by Eq.~\eqref{calc_P}.
%The phase mode for this situation is displayed in Fig.~\ref{fig:modes_dist}.
Figure~\ref{fig:modes_dist} shows that the phase mode is massive in this case, i.e., $\omega(\Q)\propto (\Q^2+C)$  for $\Q\rightarrow 0$ (with a constant $C>0$). Apart from the uniform limit, the spectrum resembles the result for the undistorted lattice since the influence of the phonons is weak for large excitation energies.

\begin{figure*}
\centering
\includegraphics[width=0.4\linewidth]{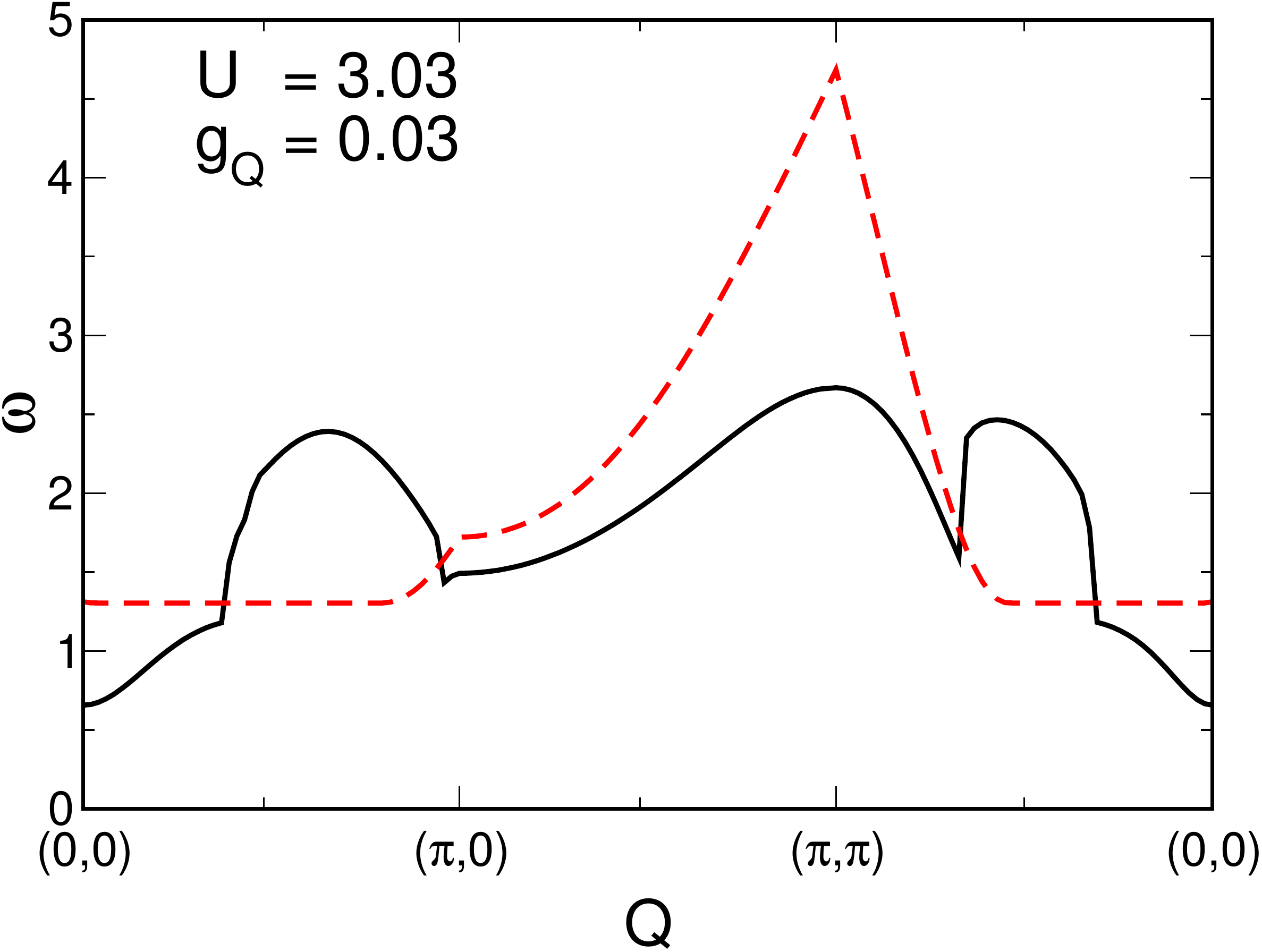}
\hspace{0.1\linewidth}
\includegraphics[width=0.4\linewidth]{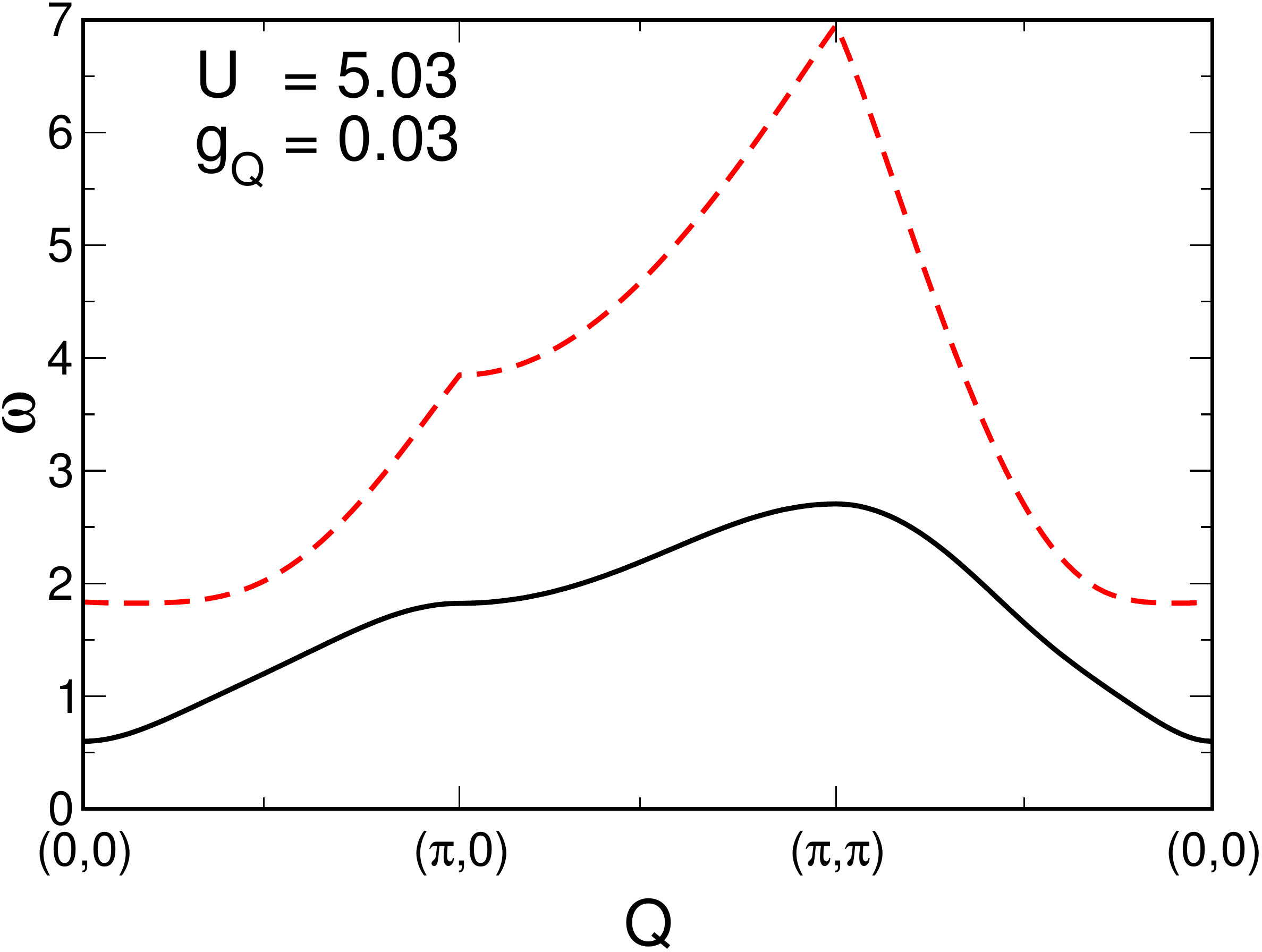}
\caption{(Color online) Electron-hole excitation spectrum for a dynamical electron-phonon coupling in instantaneous approximation at zero temperature. The black, solid lines show the phase mode and the red, dashed lines show the lower boundary of the electron-hole continuum.}
\label{fig:modes_dyn}
\end{figure*}

The absence of the acoustic phase mode can be shown analytically. The phase correlation function exhibits a pole at $z_\nu=0$ and $\Q=0$ if the denominator of Eq.~\eqref{PhasecorrSceneA} vanishes.
For  a  deformed  lattice the (Hartree-Fock)  gap equation  takes the form
\begin{equation}
0 = 1 + \left( U + 4\frac{|g_{0}|^2}{\omega_0} \right) P^{(0)}(0,0) .
\label{GapEq2}
\end{equation}
The condition for an acoustic phase mode significantly differs from Eq.~\eqref{GapEq2}. We can argue that the static lattice distortion breaks explicitly the $U(1)$ symmetry of the model and removes the phase invariance of $\Delta_{\k\bar\Q}$. As a consequence, any phase-mode excitation requires a finite energy. Hence, the phase mode is massive.

\subsection{Dynamical electron-phonon coupling}
\label{sec:Undist_CoulombPhonon}
As just has been shown, the softening of a phonon mode and the accompanying lattice deformation lead to a massive phase mode. Let us now analyze the effect of dynamical phonons that do not become soft but offer a way to transfer electrons from the valence band to the conduction band. Thereby, we include the phonons in the Bethe-Salpeter equations, Eqs.~\eqref{G2_fourtime_calc} and \eqref{F2_fourtime_calc}, and take the self-energies resulting from the coupling to the lattice in the single-particle Green functions into account.

In particular, we ask whether the phase mode in the ordered ground state is acoustic or not. To this end, we investigate  the static, uniform limit of the phase correlation function with respect to its pole structure. We note that the electron-phonon coupling leads to an effective electron-electron interaction that is nonlocal in (imaginary) time. This complicates the numerical evaluation considerably. We therefore only consider the limiting cases of slow phonons and fast phonons in comparison to the time scale of the electron transport. For these two limits, we ask whether the additional electron-phonon interaction supports electron-hole pairing or not. To this end, we analyze the phonon contribution in the gap equations taking the following bare phonon contribution into account:
\begin{equation}
D(\q,z_\nu) = -\frac{2\omega_\q}{z_\nu^2 - \omega_\q^2} .
\label{barePhononGF}
\end{equation}

First, we assume the phonons to be much slower than the electrons.
We then neglect the frequencies $z_4$ and $z_5$, which appear in the phonon Green function, in the electron-hole pair correlation functions since they only can attain small values.
In this limit the equations determine $X(\Q,z_\nu)$ and $Y(\Q,z_\nu)$, occurring in Eq.~\eqref{Phase_func}, are given in Appendix~\ref{App:slow_quantities}.
The corresponding gap equation reads
\begin{equation}
1 = \frac{1}{-i\beta} \sum_{z_1} \frac{ U R(z_1) }{ 1-|g_{0}|^2 \bar D R(z_1) } ,
\label{GapEq_slow}
\end{equation}
where
\begin{equation}
R(z_1) = \frac{1}{N}\sum_\k \frac{i}{\Omega(\k,z_1)} , 
\label{def_R} 
\end{equation}
\begin{eqnarray}
\Omega(\k,z_1) &=& \left[ z_1-\beps_{\k v}-\sigma_{vv}(\k,z_1)\right]
\left[ z_1-\beps_{\k c}-\sigma_{cc}(\k,z_1)\right]
\nonumber \\
&&- \left|\Delta_{\k\bar\Q} + \sigma_{Fv}(\k,z_1)  \right|^2 ,
\label{def_Nkz} 
\end{eqnarray}
\begin{equation}
\bar D = \frac{1}{-i\beta} \sum_{z_\mu} D(0,z_\mu) .
\label{D_bar}
\end{equation}
The $z_1$ ($z_\mu$) are fermionic (bosonic) Matsubara frequencies.  
The structure of the phase correlation function remains complicated in this case. We note, however, that the phonon contribution simply modifies the Coulomb interaction strength in  Eqs.~\eqref{appendix_eq1} and \eqref{appendix_eq2}. In the gap equation~\eqref{GapEq_slow}, on the other hand, the phonon contribution enters in a qualitatively different way. This suggests that $P(0,0)$ does not exhibit a pole and, consequently, the phase mode is massive.

In the gap equation for slow phonons, Eq.~\eqref{GapEq_slow}, we find $\bar D = -2p(\omega_0)-1<0$ with the Bose distribution function $p(x)$, and, in that 
$\frac{1}{-i\beta}\sum_{z_1} R(z_1) >0$, we can conclude that the local Coulomb potential is weakened. Self-evidently slow phonons give rise to retardation effects 
and thereby induce  an effective long-ranged electron-hole interaction potential that reduces the effect of the local Coulomb attraction.

In the opposite limit,  when the phonons are much faster than the electrons, we can integrate out, in principle, the lattice degrees of freedom (instantaneous approximation). Considering  this limit is technical rather than physically motivated since in most materials the phonon frequency is much smaller than the characteristic electronic energy scale. Due to the fact that the qualitative behavior of the phase mode is mainly determined by the underlying symmetry of the state,  the instantaneous approximation is nevertheless instructive. In this limit, we can replace the phonon Green function according to $D(\q,\tau-\tau')=D(\q,0) \delta(\tau-\tau')$. Then, the phase correlation function in the static, uniform limit becomes
\begin{equation}
P(0,0) = \frac{P^{(0)}(0,0)}{1+\left[U-|g_0|^2 D(0,0)\right] P^{(0)}(0,0)} ,
\label{PhaseMode_instant}
\end{equation}
and the gap equation is given by
\begin{equation}
1 = \left[U+|g_0|^2 D(0,0)\right] \frac{1}{-i\beta} \sum_{z_1} R(z_1) 
\label{GapEq_instant}
\end{equation}
(again $z_1$ are fermionic Matsubara frequencies).
Obviously, the instantaneous phonons lead to  a static renormalization of the Coulomb interaction. However, in the phase correlation function~\eqref{PhaseMode_instant} the phonon contribution $|g_0|^2 D(0,0)$ enters with a negative sign, while $|g_0|^2 D(0,0)$ enters with a positive sign in the gap equation~\eqref{GapEq_instant}. This discrepancy rules out that $P(0,z_\nu)$ exhibits a pole at $z_\nu=0$. Consequently the phase mode is massive, see Fig.~\ref{fig:modes_dyn}.

Obviously, in this limit, there are no retardation effects at all, and, due to the fact that $D(0,0)=2/\omega_0>0$, the phonons enhance the strength of the local Coulomb interaction [cf. Eq.~\eqref{GapEq_instant}].

That is, if the lattice is not deformed statically the phonons affect the electrons in two ways: They enhance the effective masses of the electrons and the holes (thereby modifying the band structure)  and renormalize the Coulomb interaction. The former effect is less important for the basic mechanism of exciton condensation. The latter effect, on the other hand,  is crucial, since it generates an effective electron-electron interaction that explicitly breaks the $U(1)$ symmetry. This is demonstrated by the diagrams shown in Fig.~\ref{fig:diagrams}. Here, the incoming and outgoing branches at the vertices, i.e., at $\tau$ and $\tau'$, describe the effective two-particle interaction. For the Coulomb interaction, diagramed in the second row of Fig.~\ref{fig:diagrams}, there is one incoming and outgoing branch for the valence electrons (labeled with $v$ and $v^\dagger$, respectively) and one incoming and outgoing branch for the conduction electrons (labeled with $c$ and $c^\dagger$, respectively). Hence, the interaction $V_{\rm Coul} \propto c_{\k_1 c}^\dagger c_{\k_2 c}^\pdagger c_{\k_3 v}^\dagger c_{\k_4 v}^\pdagger$. In the ladder terms arising from the electron-phonon coupling (fourth row in Fig.~\ref{fig:diagrams}) there are two incoming branches of conduction electrons and two outgoing branches of valence electrons (or vice versa), which establish an effective electron-electron interaction
\begin{equation}
V_{\rm ph} \propto c_{\k_1 c}^\dagger c_{\k_2 v}^\pdagger c_{\k_3 c}^\dagger c_{\k_4 v}^\pdagger + c_{\k_1 v}^\dagger c_{\k_2 c}^\pdagger c_{\k_3 v}^\dagger c_{\k_4 c}^\pdagger
\propto \cos(2\phi) .
\label{form_phonInt}
\end{equation}
Here, $\phi$ denotes the phase of $\Delta_{\k\bar\Q}$.
An electron-electron interaction of identical form might appear if exchange terms are considered.~\cite{GK72,*GK73,LZ96} Such an interaction fixes $\phi$ and, consequently, destroys the acoustic phase mode.

Let us note that if the electron-phonon interaction is neglected, and the Coulomb interaction is of the form~\eqref{H_ee}, the free energy is independent of $\phi$, which leads to a gapless electron-hole excitation spectrum.~\cite{JRK67} 
Without loss of generality the order parameter $\Delta_{\k\bar\Q}$  can then be assumed to be real.~\cite{ZFB10, ZFBMB13}
Taking the electron-phonon interaction into account, a possible static lattice distortion (but also the coupling of electrons and holes to dynamical phonons without lattice dimerization) induces a phase fixation and, therefore, give rise to a massive phase mode. Of course, a more complicated form of the electron-electron interaction may also lead to a gapped electron-hole excitation spectrum. The phase $\phi$ is determined by the extremal free energy varying $\phi$ (in this regard the case of a static lattice distortion has been analyzed in Ref.~\onlinecite{ZFBMB13}). 
If (the momentum-space quantity) $\delta_{\bar\Q}$ is real, the phase of $\Delta_{\k\bar\Q}$ is pinned to zero or $\pi$, i.e., both $\Delta_{\k\bar\Q}$ and the gap parameter $x_{\k\bar\Q}$ are  real. A dynamical electron-phonon interaction does not necessarily fixate the phase of $\Delta_{\k\bar\Q}$ to zero or $\pi$, accordingly $\Delta_{\k\bar\Q}$ and $x_{\k\bar\Q}$ are, in general, complex numbers.

The phase stiffness is obtained from the second derivative of the free energy with respect to $\phi$. It corresponds to the phase-mode excitation energy for $\Q=0$. That is, $\omega(0)$ can be taken as a measure for the phase fixation.

\section{Discussion of off-diagonal long-range order of electron-hole pairs}
\label{sec:condensate}
The EI is a promising candidate to observe a BCS-BEC crossover in an equilibrium situation.\cite{BF06,PBF10,ZIBF12} Since both BCS-type superconductors and Bose-Einstein condensates exhibit off-diagonal long-range order (ODLRO),~\cite{Pe51,PO56,Ya62} the question whether the EI ground state shows ODLRO or not is obvious. Here we follow (in form)  the treatment of ODLRO for BCS superconductors (see Annett's textbook~\onlinecite{An04}, Chap.~5.7), and test  possible ODLRO for electron-hole pairs.\cite{HH74}

The one-particle density matrix for bound electron-hole pairs $\rho_1^{\rm X}({\bf R}-{\bf R}')$ is related to the two-particle density matrix for electrons and holes by
\begin{equation}
\rho_1^{\rm X}({\bf R}-{\bf R}') = \int d{\bf r} \int d{\bf r}' \Psi({\bf r}) \Psi({\bf r}')
\rho_2^{\rm e-h} ({\bf r},{\bf r}',{\bf R},{\bf R}'),
\label{singleDM_twoDM}
\end{equation} 
where ${\bf R}$ and ${\bf R}'$ denote the center-of-mass coordinates of the excitons, ${\bf r}$ and ${\bf r}'$ are the relative coordinates of the (bound) electron and hole in the exciton, respectively, and $\Psi({\bf r})$ denotes the excitonic wave function. The two-particle density matrix for electrons and holes in Eq.~\eqref{singleDM_twoDM} is given by
\begin{align}
\rho&_2^{\rm e-h} ({\bf r},{\bf r}',{\bf R},{\bf R}') = 
\nonumber \\
&\frac{1}{N} \left\langle c_{c}^\dagger ({\bf R}+{\bf r}/2)
c_{v}^\pdagger ({\bf R}-{\bf r}/2) c_{v}^\dagger ({\bf R'}+{\bf r'}/2)
c_{c}^\pdagger ({\bf R'}-{\bf r'}/2) \right\rangle .
\label{densMatrix}
\end{align}
ODLRO is present if the one-particle density matrix for electron-hole pairs $\rho_1^{\rm X}({\bf R}-{\bf R}')$ remains finite for arbitrarily large separated pairs. That is, $\rho_2^{\rm e-h} ({\bf r},{\bf r}',{\bf R},{\bf R}')$ [Eq.~\eqref{densMatrix}] stays finite for $|{\bf R}-{\bf R}'|\rightarrow \infty$.

Fourier  transformation of $\rho_2^{\rm e-h}$ yields
\begin{eqnarray}
\rho_2^{\rm e-h} &=& \frac{1}{N^2} \sum_{\k,\k',\q} \langle c_{\k+\q/2 \,c}^\dagger 
c_{\k-\q/2 \,v}^\pdagger c_{\k'-\q/2 \,v}^\dagger c_{\k'+\q/2\, c}^\pdagger \rangle 
\nonumber \\
&&\times e^{i\k{\bf r}} e^{i\k'{\bf r}'} e^{i\q({\bf R}-{\bf R}')} .
\label{rho_FT}
\end{eqnarray}
At this point we stop in following Ref.~\onlinecite{An04} because the order parameter $\Delta_{\k\bar\Q}$ gives no deeper insights into the nature of the excitonic ground state. $\Delta_{\k\bar\Q}$ is finite for low temperatures regardless of the specific mechanisms which drive the phase transition and establish long-range order (BCS-type electron-hole pairing or condensation of tightly bound, preformed excitons). This is different from BCS superconductors and Bose-Einstein condensates, where the (mean-field) order parameters unambiguously characterize superconductivity, respectively, superfluidity. That is, a decoupling of Eq.~\eqref{rho_FT}, that assigns $\rho_2^{\rm e-h}$ with the order parameter $\Delta_{\k\bar\Q}$, would be a too crude approximation in our case. Therefore we relate the density matrix to  the pair correlation functions which contain valuable information about the forces driving the electron-hole pairing and condensation process.

The extent of the excitons, given by $|{\bf r}|$ and $|{\bf r}'|$, are of the order of the electron-hole pair coherence length, which is  small compared with the system size. We therefore neglect the ${\bf r}$- and ${\bf r}'$-dependencies in the following and write
\begin{eqnarray}
\rho_2^{\rm e-h} &=&-\frac{1}{N\beta} \sum_\q \sum_{z_\nu} X(\q,z_\nu) 
e^{i\q({\bf R}-{\bf R}')}
\nonumber \\
&=&-\sum_\q e^{i\q({\bf R}-{\bf R}')} I_\q ,
\label{rho_calc}
\end{eqnarray}
with $X(\q,z_\nu)=\frac{i}{N}\sum_{\k,\k'} G_2(\q,\k,\k',z_\nu)$ ($z_\nu$ are bosonic Matsubara frequencies).
The condition for ODLRO can only be satisfied if $\rho_2^{\rm e-h}$ contains averages $I_q$ of the order of unity.~\cite{Ya62}

Since we have found only one pole for a given momentum  in our numerics, in what follows we restrict ourselves to the case that $X(\q,z_\nu)$ 
exhibits a single pole (the generalization to multiple poles would be straightforward). We have
\begin{equation}
I_\q = \frac{1}{N\beta} \sum_{z_\nu} X(\q,z_\nu) = \frac{1}{N} R(\q,\omega_X) ,
\label{I_q}
\end{equation}
where $R(\q,\omega_X)$ is the residuum of the pole $\omega_X$ at momentum $\q$. For sufficiently low-lying poles we find $R(\q,\omega_X) \propto p(\omega_X)$ (note that the boundary to the electron-hole continuum is located at finite energies).\cite{ZIBF12} For  $R(\omega_X)$ to be of the order of $N$, $\omega_X$ must vanish. That is, the presence of ODLRO of electron-hole pairs implies a gapless electron-hole excitation spectrum.

Since any finite electron-phonon coupling introduces a gap, in our model ODLRO is only present if the EI phase transition is driven by the electronic correlations caused by the Coulomb interaction of type Eq.~\eqref{H_ee}.

Regardless of the particular driving mechanism, $\Delta_{\k\bar \Q}$ serves as an order parameter for the low-temperature long-range ordered phase. 
Phase fluctuations of $\Delta_{\k\bar\Q}$ may destroy the ordered state, e.g., in one-dimensional systems or two-dimensional systems at finite temperatures.\cite{Ho67} The lattice degrees of freedom may suppress these fluctuations, supporting thereby long-range order. 
In this connection, we like to emphasize that the nature of the ordered low-temperature phase in the purely electronic model, exhibiting a $U(1)$ symmetry in the normal phase, significantly differs  from the low-temperature phase in the model containing the coupling to the lattice, where the $U(1)$ symmetry is absent even in the normal phase. For the latter, ODLRO is absent (see discussion above), and we therefore suppose that a finite $\Delta_{\k\bar\Q}$ is not indicative of  any kind of  ``electron-hole pair condensate" with ``supertransport" properties. To date the identification of a measurable quantity verifying ODLRO in the materials considered as potential candidates for realizing the EI phase is, to the best of our knowledge, an open problem.

\section{Conclusions}
\label{sec:conclusion}
In this work we have revisited on what terms an excitonic insulator (EI) forms. In particular, we have analyzed the effects of an explicit electron-phonon interaction $H_{\rm e-ph}$. The potential EI state then may possess a static lattice distortion. We have shown that $H_{\rm e-ph}$ will not change the single-particle spectra qualitatively, even if self-energy effects are taken into account. However,  $H_{\rm e-ph}$ significantly modifies the electron-hole pair spectrum. To demonstrate this, we have calculated the contributions of the electron-phonon interaction to  electron-hole pairing within the Kadanoff-Baym approach including ring and ladder diagrams. When the electron-phonon coupling is neglected the phase mode is acoustic. Electron-lattice coupling destroys the acoustic mode regardless if it causes a static lattice distortion or renormalizes the effective electron-electron interaction.

We pointed out that an acoustic phase mode implies the presence of off-diagonal long range order (ODLRO), and therefore indicates---in a strongly coupled electron-hole system---an exciton ``condensate". This applies to the EI phase in pure electronic models as, e.g., the extended Falicov-Kimball model.~\cite{Ba02b, BGBL04, IPBBF08,ZIBF12}  Since in most of the (potential) EI  materials considered so far, the lattice degrees of freedom play a non-negligible role, they should prevent, according to  the reasonings of this paper, the appearance of an acoustic phase mode. Hence these materials embody rather unusual (gapped) charge-density-wave systems than true exciton condensates with super-transport properties (cf. the remark by Kohn in the supplementary discussion in Ref.~\onlinecite{HR68a}).

To realize an  exciton condensate in equilibrium experimentally, bilayer systems, such as graphene double layers and bilayers,~\cite{LY76, LY77,LS08,DH08,MBSM08,KE08,PF12,PNH13} are the most promising candidates at present. Since the interband tunneling processes can be suppressed by suitable dielectrics, an acoustic collective mode, and hence ODLRO, may emerge.
In these systems electrons and holes occupy different layers and the exciton condensation is presumably accompanied by the appearance of supercurrents in both layers that flow in opposite directions,~\cite{LY76} respectively, the occurrence of a dipolar supercurrent.~\cite{BJL04}

Let us finally emphasize that the numerical results presented in this work are obtained using rather crude approximations. That is why a more elaborated numerical treatment is highly desirable.
A possible next step is to calculate the dynamical structure factor, which is accessible experimentally  by electron energy-loss spectroscopy.~\cite{WSKKBBB11} Here, collective modes show up as peaks and one might address the acoustic phase-mode problem.
The phase invariance leading to the acoustic phase mode might also be reflected in Josephson-like phenomena induced by tunneling excitons.
Moreover, the behavior of the plasmon mode in the low-temperature state has not been elaborated yet. This mode is generated by intraband correlations and shows an acoustic behavior in the normal phase.~\cite{GU73} 
We mentioned that the inclusion of exchange terms in the Coulomb interaction destroys the phase invariance, just as the electron-phonon interaction considered in this work.~\cite{GK73, LZ96}  However, electron-electron and electron-phonon interactions do not have to promote the same values for the phase of $\Delta_{\k\bar\Q}$; thus it is interesting to analyze the consequences of their interplay. In particular,  cooling down the system, the phase realized may alter.
Another worthwhile continuation concerns the possible formation and condensation of ``polaron excitons," i.e., the buildup of a condensate of excitons which are dressed by a phonon cloud.

\section*{Acknowledgements}
This work was supported by the DFG through SFB 652, project B5. 

%\bibliography{paper} 
\bibliographystyle{apsrev}

\appendix
\section{Equations of motion for the single-particle Green functions}
\label{app:EOMs}
The equation of motion (EOM) for the conduction-electron Green function is given by
\begin{align}
\bigg( i\frac{\partial}{\partial t}- & \beps_{\k c} \bigg) G_c(\k,t-t')
=\delta(t-t') + x_{\k\bar\Q} F^\dagger(\k-\bar\Q,t-t') 
\nonumber \\
&-\int_0^{-i\beta} d\!\tau\, \sigma_{cc}(\k,t-\tau) G_c(\k,\tau-t')
\nonumber \\
&- \int_0^{-i\beta} d\!\tau\, \sigma_{cF}(\k,t-\tau) F^\dagger(\k-\bar\Q,\tau-t') 
\label{EOM_Gc_final}
\end{align}
with
\begin{align}
\sigma_{cc}(\k&,t-\tau) = \frac{1}{N^2} \sum_{\q,\q',\Q} V_c(\q) V_c(\q') G_v(\k-\Q,t-\tau)
\nonumber \\
&\times G_2(\Q,\k+\q-\Q,\k-\q'-\Q,t-\tau)
\nonumber \\
&-\frac{i}{N}\sum_\q |g_\q|^2 D(\q,t-\tau) G_v(\k-\q,t-\tau) ,
\label{sigma_cc}
\end{align}
\begin{align}
\sigma_{cF}(\k&,t-\tau) = \frac{1}{N^2}\sum_{\q,\q',\Q} V_c(\q) V_c(\q')
F^\dagger(\k-\Q,t-\tau)
\nonumber \\
&\times H_2(\Q,\k+\q'-\bar\Q,\k+\q-\Q,\tau-t)
\nonumber \\
&-\frac{i}{N}\sum_\q |g_\q|^2 D(\q,\tau-t) F^\dagger(\k+\q,t-\tau) ,
\label{sigma_cF}
\end{align}
and
\begin{equation}
H_2(\Q,\k,\k',t-t') = -\left\langle T[ c_{\k v}^\dagger(t) c_{\k-\Q c}^\pdagger(t)
c_{\k' v}^\dagger (t') c_{\k'+\Q c}^\pdagger (t') ] \right\rangle .
\label{H2}
\end{equation}

The EOM for the anomalous Green function reads
\begin{align}
\bigg( i\frac{\partial}{\partial t} - & \beps_{\k+\bar\Q c} \bigg) F(\k,t-t')
= x_{\k\bar\Q} G_v(\k,t-t') 
\nonumber \\
&-\int_0^{-i\beta} d\!\tau\, \sigma_{Fv}(\k,t-\tau) G_v(\k,\tau-t')
\nonumber \\
&- \int_0^{-i\beta} d\!\tau\, \sigma_{FF}(\k,t-\tau) F(\k,\tau-t') 
\label{EOM_F_final}
\end{align}
with
\begin{align}
\sigma&_{Fv} (\k,t-\tau) = \frac{1}{N^2} \sum_{\q,\q',\Q} V_c(\q) V_c(\q') F(\k+\bar\Q-\Q,t-\tau) 
\nonumber \\
&\times H_2(\Q,\k+\q',\k+\q+\bar\Q-\Q,\tau-t)
\nonumber \\
&-\frac{i}{N}\sum_\q |g_\q|^2 D(\q,t-\tau) F^\dagger(\k+\bar\Q-\q,t-\tau) ,
\label{sigma_Fv} 
\end{align}
\begin{align}
\sigma&_{FF} (\k,t-\tau) = \frac{1}{N^2}\sum_{\q,\q',\Q} V_c(\q) V_c(\q') G_v(\k+\bar\Q-\Q,t-\tau) 
\nonumber \\
&\times G_2(\Q,\k+\q+\bar\Q,\k+\bar\Q-\Q-\q',t-\tau)
\nonumber \\
&-\frac{i}{N}\sum_\q |g_\q|^2 D(\q,t-\tau) G_v(\k+\bar\Q-\q,t-\tau) .
\label{sigma_FF}
\end{align}

\begin{widetext}
\section{Equations for the electron-hole pair correlation functions}
\label{app:correlationFuncs}
If both Coulomb and phonon effects  are of importance, the electron-hole pair correlation function~\eqref{G2_fourtime} has to be calculated according to
\begin{align}
G_2(\Q,\k, & \k',t_1,t_2,t_3,t_4) = G_2^{(0)}(\Q,\k,t_1,t_2,t_3,t_4) \delta_{\k,\k'}
\nonumber \\
&-\frac{i}{N}\int_0^{-i\beta} d(\tau-t_4) \sum_\q V_c(\q) G_2^{(0)}(\Q,\k,t_1,t_2,\tau,\tau)  
G_2(\Q,\k+\q,\k',\tau,\tau,t_3,t_4)
\nonumber \\
&-\frac{i}{N}\int_0^{-i\beta} d(\tau-t_4) \sum_\q V_c(\q) F_2^{(0)}(\Q,\k,t_1,t_2,\tau,\tau)  
F_2(\Q,\k+\q+\Q+\bar\Q,\k',\tau,\tau,t_3,t_4)
\nonumber \\
&+\frac{i}{N}\int_0^{-i\beta} d(\tau-t_4) 
\int_0^{-i\beta}d(\tau'-t_4) \sum_{\q} \left( |g_\Q|^2 D(\Q,\tau-\tau') 
[ G_2^{(0)}(\Q,\k,t_1,t_2,\tau,\tau) \right.
\nonumber \\
&+ F_2^{(0)}(\Q,\k,t_1,t_2,\tau,\tau)]
[ G_2(\Q,\q,\k',\tau',\tau',t_3,t_4) + F_2(\Q,\q,\k',\tau',\tau',t_3,t_4)]
\nonumber \\
&-|g_{\Q+\k-\q}|^2 D(\Q+\k-\q,\tau'-\tau) G_2^{(0)}(\Q,\k,t_1,t_2,\tau,\tau')
F_2(\Q,\q,\k',\tau,\tau',t_3,t_4)
\nonumber \\
&\left. -|g_{\k-\q}|^2 D(\k-\q,\tau'-\tau) F_2^{(0)}(\Q,\k,t_1,t_2,\tau,\tau')
G_2(\Q,\q,\k',\tau,\tau',t_3,t_4) \right) .
\label{G2_fourtime_calc}
\end{align}
%\end{widetext}
where the phonon Green function is defined by Eq.~\eqref{D}, and
\begin{equation}
G_2^{(0)}(\Q,\k,t_1,t_2,t_3,t_4) = -G_v(\k,t_3-t_1) G_c(\k+\Q,t_2-t_4) ,
\label{G2_0} 
\end{equation}
\begin{equation}
F_2^{(0)}(\Q,\k,t_1,t_2,t_3,t_4) = -F(\k,t_3-t_1) F(\k+\Q,t_2-t_4) .
\label{F2_0}
\end{equation}
Typical diagrams occurring in Eq.~\eqref{G2_fourtime_calc} are shown in Fig.~\ref{fig:diagrams}.

If $x_{\k\bar\Q}\neq0$, $G_2(\Q,\k,\k',t_1,t_2,t_3,t_4)$ is coupled to $F_2(\Q,\k,\k',t_1,t_2,t_3,t_4)$. $F_2$  can be determined from
\begin{align}
F_2(\Q,\k, & \k',t_1,t_2,t_3,t_4) = \bar F_2^{(0)}(\Q,\k,t_1,t_2,t_3,t_4) 
\delta_{\k+\bar\Q,\k'+\Q}
\nonumber \\
&-\frac{i}{N}\int_0^{-i\beta} d(\tau-t_4) \sum_\q V_c(\q) 
\bar F_2^{(0)}(\Q,\k,t_1,t_2,\tau,\tau)  
G_2(\Q,\k+\q+\bar\Q-\Q,\k',\tau,\tau,t_3,t_4)
\nonumber \\
&-\frac{i}{N}\int_0^{-i\beta} d(\tau-t_4) \sum_\q V_c(\q) 
\bar G_2^{(0)}(\Q,\k,t_1,t_2,\tau,\tau)  
F_2(\Q,\k+\q,\k',\tau,\tau,t_3,t_4)
\nonumber \\
&+\frac{i}{N}\int_0^{-i\beta} d(\tau-t_4) 
\int_0^{-i\beta}d(\tau'-t_4) \sum_{\q} \left( |g_\Q|^2 D(\Q,\tau-\tau') 
[ \bar G_2^{(0)}(\Q,\k,t_1,t_2,\tau,\tau)  \right.
\nonumber \\
&+ \bar F_2^{(0)}(\Q,\k,t_1,t_2,\tau,\tau)]
[ G_2(\Q,\q,\k',\tau',\tau',t_3,t_4) + F_2(\Q,\q,\k',\tau',\tau',t_3,t_4)]
\nonumber \\
&-|g_{\Q+\q-\k}|^2 D(\Q+\q-\k,\tau-\tau') \bar G_2^{(0)}(\Q,\k,t_1,t_2,\tau,\tau')
G_2(\Q,\q,\k',\tau,\tau',t_3,t_4)
\nonumber \\
&\left. -|g_{\q-\k}|^2 D(\q-\k,\tau-\tau') \bar F_2^{(0)}(\Q,\k,t_1,t_2,\tau,\tau')
F_2(\Q,\q,\k',\tau,\tau',t_3,t_4) \right) ,
\label{F2_fourtime_calc}
\end{align}
where
\begin{equation}
\bar G_2^{(0)}(\Q,\k,t_1,t_2,t_3,t_4) = -G_c(\k-\Q,t_3-t_1) G_v(\k,t_2-t_4) ,
\label{barG2_0} 
\end{equation}
\begin{equation}
\bar F_2^{(0)} (\Q,\k,t_1,t_2,t_3,t_4) = -F^\dagger(\k-\Q,t_3-t_1) F^\dagger(\k,t_2-t_4) .
\label{barF2_0}
\end{equation}

For an explicit calculation of the electron-hole pair correlation functions the Matsubara technique is  advantageous. Performing the transformation
\begin{align}
G_2 (\Q,\k,\k',z_1,z_2,z_3) =& \int_0^{-i\beta} d(t_1-t_4) e^{-i z_1(t_1-t_4)}
\int_0^{-i\beta} d(t_2-t_4) e^{-i z_2(t_2-t_4)}
\int_0^{-i\beta} d(t_3-t_4) e^{-i z_3(t_3-t_4)}
\nonumber \\
&\times G_2(\Q,\k,\k',t_1,t_2,t_3,t_4) ,
\label{FourierTrafo}
\end{align}
we obtain
\begin{align}
G_2&(\Q,\k,\k',z_{1},z_{2},z_{3}) 
= G_2^{(0)}(\Q,\k,z_{1},z_{2},z_{3}) \delta_{\k,\k'}
\nonumber \\
&-\frac{i}{N}\left(\frac{1}{-i\beta}\right)^2 \sum_{z_4,z_5} \sum_\q V_c(\q)
G_2^{(0)}(\Q,\k,z_{1},z_{2},z_4) G_2(\Q,\k+\q,\k',z_5,z_{1} + 
z_{2}-z_5,z_{3})
\nonumber \\
&-\frac{i}{N}\left(\frac{1}{-i\beta}\right)^2 \sum_{z_4,z_5} \sum_\q V_c(\q)
F_2^{(0)}(\Q,\k,z_{1},z_{2},z_4) F_2(\Q,\k+\q+\Q+\bar\Q,\k',z_5,z_{1} + z_2-z_5,z_{3})
\nonumber \\
&+\frac{i}{N}\left(\frac{1}{-i\beta}\right)^2 \sum_{z_4,z_5} \sum_\q
|g_\Q|^2 D(\Q,z_{1}+z_{2}) \left[ G_2^{(0)}(\Q,\k,z_{1},z_{2},z_4)
+ F_2^{(0)}(\Q,\k,z_{1},z_{2},z_4) \right]
\nonumber\\
&\times \left[ G_2(\Q,\q,\k',z_5,z_{1}+z_{2}-z_5,z_{3})
+ F_2(\Q,\q,\k',z_5,z_{1}+z_{2}-z_5,z_{3})\right]
\nonumber \\
&-\frac{i}{N}\left(\frac{1}{-i\beta}\right)^2 \sum_{z_4,z_5} \sum_\q
\left[ |g_{\Q+\k-\q}|^2 D(\Q+\k-\q,z_4+z_5) \right.
\nonumber \\
& \times G_2^{(0)}(\Q,\k,z_{1},z_{2},z_4) 
F_2(\Q,\q,\k',z_5,z_{1}+z_{2}-z_5, z_{3}) 
\nonumber \\
&\left. + |g_{\k-\q}|^2 D(\k-\q,z_4+z_5)
F_2^{(0)}(\Q,\k,z_{1},z_{2},z_4)
G_2(\Q,\k-\q,\k',z_5,z_{1}+z_{2}-z_5, z_{3}) \right]
\label{G2_threeFreq}
\end{align}
and
\begin{align}
F_2&(\Q,\k,\k',z_{1},z_{2},z_{3}) 
= \bar F_2^{(0)}(\Q,\k,z_{1},z_{2},z_{3}) \delta_{\k+\bar\Q,\k'+\Q}
\nonumber \\
&-\frac{i}{N}\left(\frac{1}{-i\beta}\right)^2 \sum_{z_4,z_5} \sum_\q V_c(\q)
\bar G_2^{(0)}(\Q,\k,z_{1},z_{2},z_4) F_2(\Q,\k+\q,\k',z_5,z_{1} + z_{2}-z_5,z_{3})
\nonumber \\
&-\frac{i}{N}\left(\frac{1}{-i\beta}\right)^2 \sum_{z_4,z_5} \sum_\q V_c(\q)
\bar F_2^{(0)}(\Q,\k,z_{1},z_{2},z_4) G_2(\Q,\k+\q+\bar\Q-\Q,\k',z_5,z_{1}+z_{2}-z_5,z_{3})
\nonumber \\
&+\frac{i}{N}\left(\frac{1}{-i\beta}\right)^2 \sum_{z_4,z_5} \sum_\q
|g_\Q|^2 D(\Q,z_{1}+z_{2}) \left[ \bar G_2^{(0)}(\Q,\k,z_{1},z_{2},z_4)
+\bar F_2^{(0)}(\Q,\k,z_{1},z_{2},z_4) \right]
\nonumber\\
&\times \left[ G_2(\Q,\q,\k',z_5,z_{1}+z_{2}-z_5,z_{3})
+ F_2(\Q,\q,\k',z_5,z_{1}+z_{2}-z_5,z_{3})\right]
\nonumber \\
&-\frac{i}{N}\left(\frac{1}{-i\beta}\right)^2 \sum_{z_4,z_5} \sum_\q
\left[ |g_{\Q+\q-\k}|^2 D(\Q+\q-\k,z_4+z_5) \right.
\nonumber \\
& \times \bar G_2^{(0)}(\Q,\k,z_{1},z_{2},z_4) 
G_2(\Q,\q,\k',z_5,z_{1}+z_{2}-z_5, z_{3}) 
\nonumber \\
&\left. + |g_{\q-\k}|^2 D(\q-\k,z_4+z_5)
 \bar F_2^{(0)}(\Q,\k,z_{1},z_{2},z_4)
F_2(\Q,\k-\q,\k',z_5,z_{1}+z_{2}-z_5, z_{3}) \right] ,
\label{F2_threeFreq}
\end{align}
where the $z_i$, $i=1\ldots5$, are fermionic Matsubara frequencies.

\section{Functions appearing in the phase correlation function for slow phonons}
\label{App:slow_quantities}
\begin{equation}
\tilde X^{(0)}(\Q,z_\nu)  =\left[1-r_x(\Q,z_\nu)\right] X(\Q,z_\nu) 
- r_y(\Q,z_\nu) Y(\Q,z_\nu) ,
\label{X_slow}
\end{equation}
\begin{equation}
\tilde Y^{(0)}(\Q,z_\nu)  = \left[1-s_y(\Q,z_\nu)\right] Y(\Q,z_\nu) 
- s_x(\Q,z_\nu) X(\Q,z_\nu) ,
\label{Y_slow}
\end{equation}
where
\begin{equation}
r_x(\Q,z_\nu) = \frac{1}{-i\beta} \sum_{z_2} 
\frac{A_x(\Q,z_\nu,z_2)\left[1+a(-\Q,-z_\nu,z_2)\right] 
- b(\Q,z_\nu,z_2) A_y(-\Q,-z_\nu,z_2)}
{\left[1+a(\Q,z_\nu,z_2)\right]\left[1+a(-\Q,-z_\nu,z_2)\right]
- b(\Q,z_\nu,z_2) b(-\Q,-z_\nu,z_2)} ,
\end{equation}
\begin{equation}
r_y(\Q,z_\nu) = \frac{1}{-i\beta} \sum_{z_2} 
\frac{A_y(\Q,z_\nu,z_2)\left[1+a(-\Q,-z_\nu,z_2)\right] 
- b(\Q,z_\nu,z_2) A_x(-\Q,-z_\nu,z_2)}
{\left[1+a(\Q,z_\nu,z_2)\right]\left[1+a(-\Q,-z_\nu,z_2)\right]
- b(\Q,z_\nu,z_2) b(-\Q,-z_\nu,z_2)} ,
\end{equation}
\begin{equation}
\tilde X^{(0)}(\Q,z_\nu) = \frac{1}{-i\beta} \sum_{z_2} 
\frac{\bar X^{(0)}(\Q,z_\nu,z_2)\left[1+a(-\Q,-z_\nu,z_2)\right] 
- b(\Q,z_\nu,z_2) \bar Y_2^{(0)} (-\Q,-z_\nu,z_2)}
{\left[1+a(\Q,z_\nu,z_2)\right]\left[1+a(-\Q,-z_\nu,z_2)\right]
- b(\Q,z_\nu,z_2) b(-\Q,-z_\nu,z_2)} ,
\end{equation}
\begin{equation}
s_x(\Q,z_\nu) = \frac{1}{-i\beta} \sum_{z_2} 
\frac{A_y(-\Q,-z_\nu,z_2)\left[1+a(\Q,z_\nu,z_2)\right] 
- b(-\Q,-z_\nu,z_2) A_x(\Q,z_\nu,z_2)}
{\left[1+a(\Q,z_\nu,z_2)\right]\left[1+a(-\Q,-z_\nu,z_2)\right]
- b(\Q,z_\nu,z_2) b(-\Q,-z_\nu,z_2)} ,
\end{equation}
\begin{equation}
s_y(\Q,z_\nu) = \frac{1}{-i\beta} \sum_{z_2} 
\frac{A_x(-\Q,-z_\nu,z_2)\left[1+a(\Q,z_\nu,z_2)\right] 
- b(-\Q,-z_\nu,z_2) A_y(\Q,z_\nu,z_2)}
{\left[1+a(\Q,z_\nu,z_2)\right]\left[1+a(-\Q,-z_\nu,z_2)\right]
- b(\Q,z_\nu,z_2) b(-\Q,-z_\nu,z_2)} ,
\end{equation}
\begin{equation}
\tilde Y^{(0)}(\Q,z_\nu) = \frac{1}{-i\beta} \sum_{z_2} 
\frac{\bar Y^{(0)}(-\Q,-z_\nu,z_2)\left[1+a(\Q,z_\nu,z_2)\right] 
- b(-\Q,-z_\nu,z_2) \bar X_2^{(0)} (\Q,z_\nu,z_2)}
{\left[1+a(\Q,z_\nu,z_2)\right]\left[1+a(-\Q,-z_\nu,z_2)\right]
- b(\Q,z_\nu,z_2) b(-\Q,-z_\nu,z_2)} ,
\end{equation}
\begin{equation}
A_x(\Q,z_\nu,z_2) = -U \bar X_2^{(0)}(\Q,z_\nu,z_2) + |g_0|^2 D(0,z_\nu)
\left[ \bar X_2^{(0)}(\Q,z_\nu,z_2) + \bar Y_2^{(0)}(\Q,z_\nu,z_2) \right] ,
\label{appendix_eq1}
\end{equation}
\begin{equation}
A_y(\Q,z_\nu,z_2) = -U \bar Y_2^{(0)}(\Q,z_\nu,z_2) + |g_0|^2 D(0,z_\nu)
\left[ \bar X_2^{(0)}(\Q,z_\nu,z_2) + \bar Y_2^{(0)}(\Q,z_\nu,z_2) \right] ,
\label{appendix_eq2}
\end{equation}
\begin{equation}
a(\Q,z_\nu,z_2) = |g_0|^2 \bar D(0) \bar Y_2^{(0)}(\Q,z_\nu,z_2) ,
\end{equation}
\begin{equation}
b(\Q,z_\nu,z_2) = |g_0|^2 \bar D(0) \bar X_2^{(0)}(\Q,z_\nu,z_2) ,
\end{equation}
\begin{equation}
\bar X_2^{(0)}(\Q,z_\nu,z_2) = \frac{1}{-i\beta} \frac{i}{N} \sum_{\k,\k'} \sum_{z_3} G_2^{(0)}(\Q,\k,\k',z_\nu-z_2,z_2,z_3) ,
\end{equation}
\begin{equation}
\bar Y_2^{(0)}(\Q,z_\nu,z_2) = \frac{1}{-i\beta} \frac{i}{N} \sum_{\k,\k'} \sum_{z_3} F_2^{(0)}(\Q,\k,\k',z_\nu-z_2,z_2,z_3) .
\end{equation}
\end{widetext}

\end{document}